\documentclass[review]{siamart0516}

\usepackage{lipsum}
\usepackage{amsfonts}
\usepackage{graphicx}
\usepackage{epstopdf}
\usepackage{algorithmic}
\ifpdf
  \DeclareGraphicsExtensions{.eps,.pdf,.png,.jpg}
\else
  \DeclareGraphicsExtensions{.eps}
\fi

\newcommand{\TheTitle}{A Markov Chain Surrogate Model for a Two-Dimensional Interacting Particle System with Internal Collisions} 
\newcommand{\TheAuthors}{  }

\headers{MC Surrogate Model for a 2D Interacting Particle System}{\TheAuthors}

\title{{\TheTitle}\thanks{\ }
}

\author{
  Tricity Andrew\thanks{Department of Physics, North Carolina State University, Raleigh, NC (\email{tmandre2@ncsu.edu}).}
    \and James D.~Nance\thanks{Numerica Corporation (\email{james.nance@numerica.us}).}
  \and
  Mansoor A.~Haider\thanks{Department of Mathematics, North Carolina State University, Raleigh, NC (\email{mahaider@ncsu.edu}), Corresponding author.}
}

\usepackage{amsopn}

\ifpdf
\hypersetup{
  pdftitle={\TheTitle},
  pdfauthor={\TheAuthors}
}
\fi

\usepackage{graphicx} 
\usepackage{subcaption}
\begin{document}

\maketitle

\begin{abstract}
A probabilistic Markov Chain (MC) surrogate model for a two-dimensional system of interacting particles within a square domain having inherent symmetries is developed. 
Particles are assumed to be circular and identical, colliding with each other and with rigid domain walls via perfectly elastic collisions.  
Simulation results over many realizations are used to develop and evaluate the surrogate MC model.
The stationary quantity of interest (QoI) is the number of particles within each of nine coarser subdomains, delineated via three internal geometric symmetries. 
The surrogate model is used to quantify QoI mean properties and uncertainty as the particle radius is varied.  
\end{abstract}


\section{Introduction}

In modeling material or biological systems, direct simulation is often used to investigate  interactions among system components at microscopic scales. 
For example, such an approach may be considered when the full coupling of governing equations at disparate scales is difficult due to computational complexity or the lack of appropriate theoretical models describing interactions among system variables at coarser scales  (Alarcon et al.~\cite{Alarcon}; Jacob \cite{Jacob}).  
Examples of direct simulation approaches include the use of representative domains (volume elements)  (Guo et al.~\cite{Guo};  Halloran et al.~\cite{Halloran}; Kanit et al.~\cite{Kanit}), agent-based models (An et al.~\cite{An}; Jamali et al.~\cite{Jamali}; Poleszczuk et al.~\cite{Poles}) or lattice-based models in biological applications (Anderson et al.~\cite{Anderson}; Jiang et al.~\cite{Jiang}).
These approaches have been applied to simulate and identify complex or heterogeneous interactions on microscopic scales, and have seen wide use in a variety of applications.  
Yet, there remains a need for approaches that bridge finer-scale models, based on direct simulation, to coarser-scale models, e.g. based on continuum theories. 
One approach to addressing this need is to systematically identify stationary or stochastic dynamic properties of simulation results to develop, when possible or practical, probabilistic surrogate models. 

Many interacting particle systems exhibit structure in their dynamics due to homogeneities of the particles or materials in the system, inherent geometric symmetries in the simulation domain, or a limited set of rules or natural laws governing their interactions. 
Data obtained via direct simulation over many realizations can facilitate systematic evaluation of dynamic transitions among quantities of interest.
This, in turn, can enable quantification of the mean properties of stationary system variables as well as  uncertainty due to inherent stochasticity.    
The approach considered in this study develops such a surrogate model that preserve the statistical features of key system variables using a simpler model representation  in the context of a two-dimensional model of colliding circular particles within a square domain with internal symmetries. 

In our model problem,  quantities of interest can be tracked by classification into states at each point in time, and transitions between these states at successive times; hence, Markov chains (Privault \cite{Privault}) provide a  framework for developing a surrogate model. 
 In the regime where dynamics are simulated over long time periods and  many realizations, distributions for key quantities of interest can be identified and represented via a small number of parameters, provided that appropriate probability density functions can be identified (e.g.~mean and variance  in a Gaussian distribution).  
 Overall, the success of this approach will depend on several factors.  
  First, the direct simulation model  must be robust and efficient enough to exhibit stationary statistical properties over appropriate time scales and over many realizations.  
  Second, the number and nature of the states identified for use in the Markov chain (MC) model must be such that the surrogate model can, in a practical sense, reproduce the full continuous (directly simulated) model. 
  And, lastly, the data (histograms) for stationary quantities of interest obtained using the surrogate MC model should be self-consistent in that a family of distributions can be identified to accurately reproduce them via curve-fitting. 

Based on this approach, this study develops and evaluates a probabilistic MC surrogate model in the context of a two-dimensional model problem comprising interacting circular particles within a square domain that exhibits some internal geometric symmetries.  
The particles are assumed to all have the same radius and to interact with each other, and with the rigid walls of the square domain, via perfectly elastic collisions.  
System dynamics are captured using a direct simulation model that enforces conservation of momentum for all collisions.  The stationary quantity of interest is chosen as the number of particles within each of nine coarser subdomains, delineated based on three types of geometric features.   
Direct  simulation of the  continuous model is used to develop  a surrogate (probabilistic)  Markov Chain model and evaluate its accuracy.  The resulting models are then used to quantify mean properties and  uncertainty of the stationary variable as the particle radius is varied.  

\section{Models} 
We first describe the continuous model used to directly simulate particle interactions (Sec.~\ref{sec:ctsmodel}) and then outline how the results can be used to develop a surrogate model, based on Markov chains, to accurately capture the system dynamics in an alternate manner (Sec.~\ref{sec:MCmodel}). 

\subsection{Continuous Model of Particle Interactions} \label{sec:ctsmodel}

\begin{figure}[!h]
\begin{center}
  \includegraphics[scale=0.40]{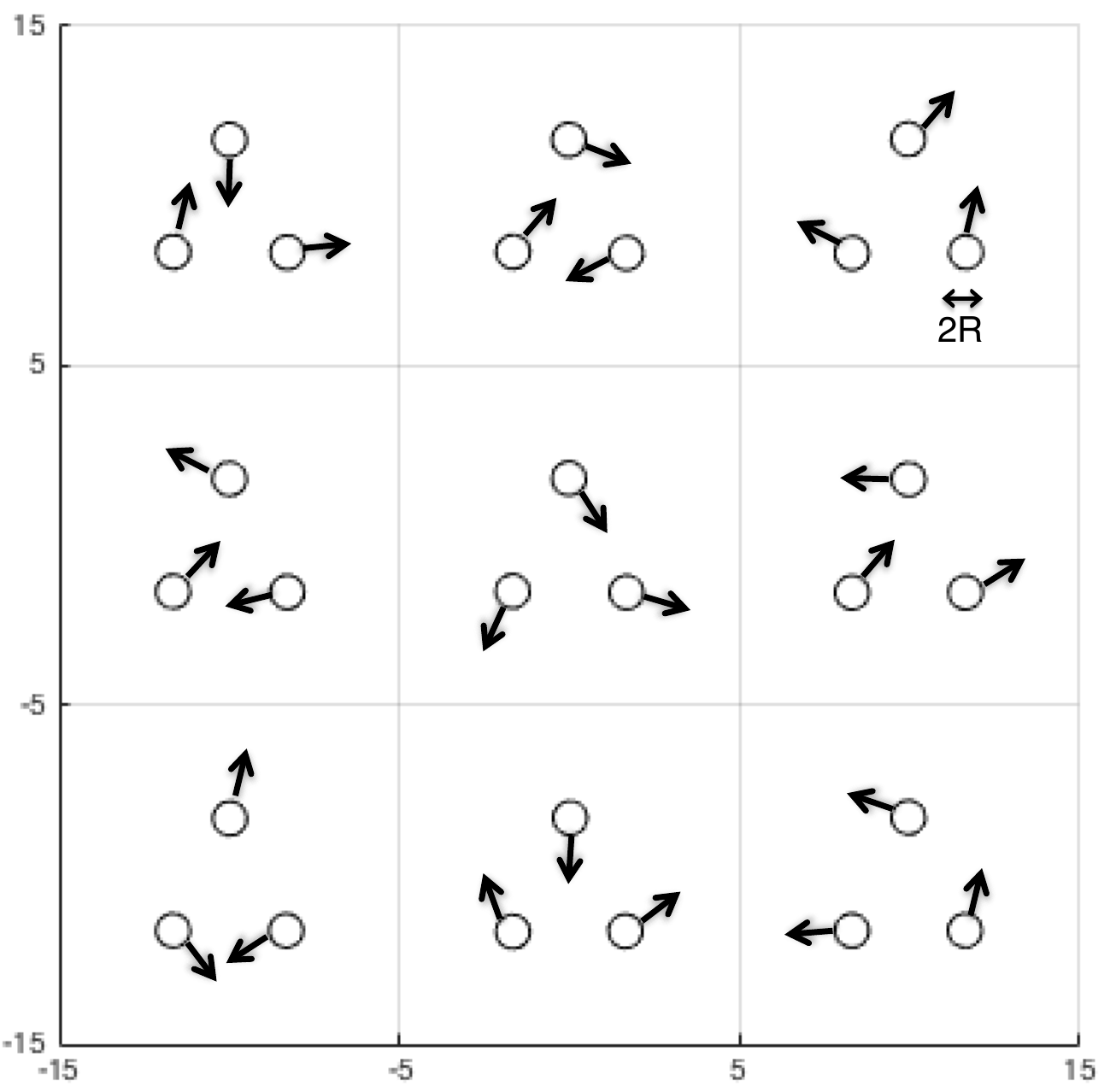} \ \ \ 
  \includegraphics[scale=0.40]{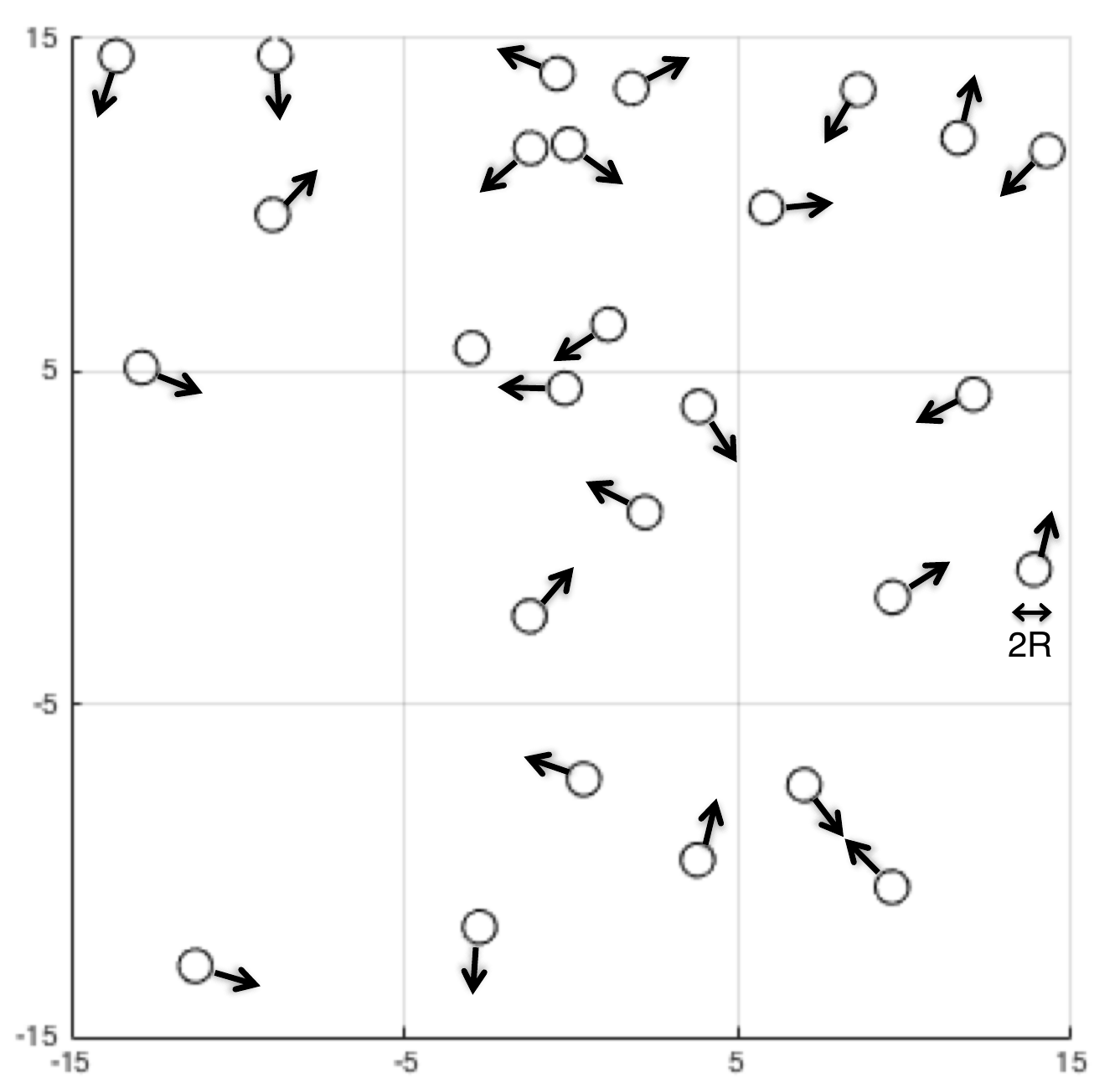}
  \caption{Two dimensional dynamic  model for perfectly elastic collisions between 27 particles and four rigid walls in a square domain: (left) the initial state ($t=0$) for a sample realization,  (right) an intermediate state ($t>0$).}
  \label{fig:Part}
\end{center}
\end{figure}

We consider a square domain in two spatial dimensions with four rigid walls. Within this domain, 27 circular particles having identical mass and radius ($R$) are undergoing motion involving dynamic collisions with the other particles and with the four walls of the domain  (Fig.~\ref{fig:Part}-right).  To facilitate development of the surrogate model (Sec.~\ref{sec:MCmodel}), the square domain is viewed as the union of nine, non-overlapping  square subdomains of equal area. At $t=0$, the 27 particles are partitioned into nine groups of three particles each and initially placed, as illustrated,    into each of the nine subdomains (Fig.~\ref{fig:Part}-left).  The initial direction of motion for each particle is chosen randomly based on a uniform distribution and prescribed a constant initial speed $|{\bf v}|$. Once the system is set into motion ($t>0$), the particles are assumed to exhibit perfectly elastic collisions with each other and with the four rigid walls of the square domain, i.e.~all  collisions conserve linear momentum. 

\begin{figure}[!h]
\begin{center}
  \includegraphics[scale=0.36]{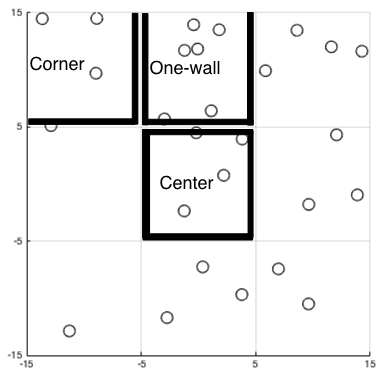} \ \ \ \ \ \ \ \ \ \ \ 
  \includegraphics[scale=0.38]{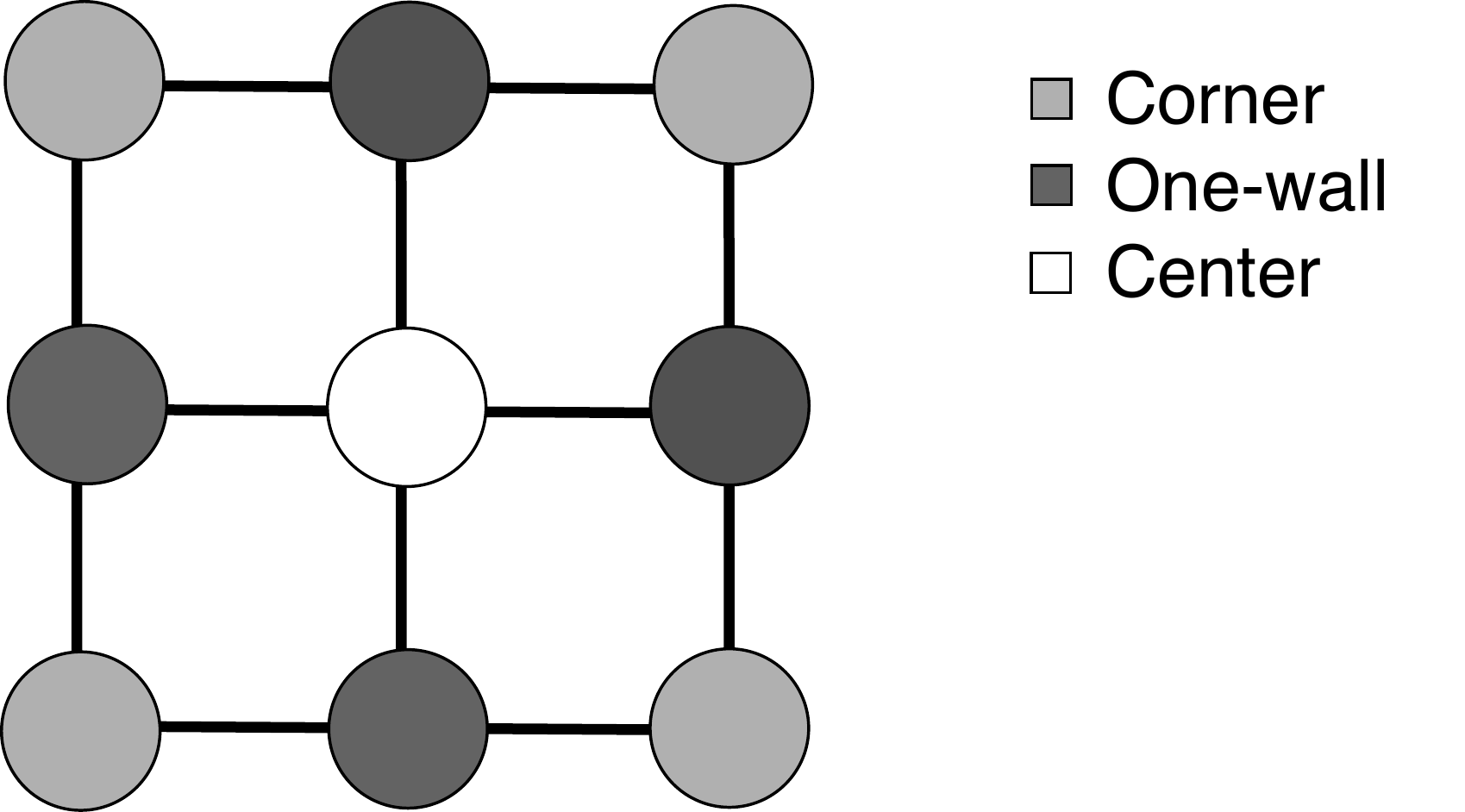}
  \caption{Partitioning of the  direct simulation domain into 9 square subdomains of equal area:  (left) subdomains are delineated into three types based on the number of subdomain boundaries, (right) illustration of adjacency relationships between the 9 subdomains.}
  \label{fig:subdomains}
\end{center}
\end{figure}

Our goal is to investigate time evolution of the number of particles per subdomain and delineate the stationary response of this quantity of interest based on the subdomain type and particle radius $R$.   The three subdomain types are referred to as {\em Corner} (two exterior walls), {\em One-wall} (one exterior wall) and {\em Center} (no exterior walls) (Fig.~\ref{fig:subdomains}-left).  The entire simulation domain thus consists of four Corner subdomains, four One-wall subdomains, and one Center subdomain, along with adjacency relationships (Fig.~\ref{fig:subdomains}-right).   

Based on the system characteristics described above, we can make the following observations:
\begin{enumerate}
\item As the particle radius $R\rightarrow 0$, the expected value of the number of particles per subdomain in the stationary response should tend to 3.
\item As the particle radius $R$ increases above zero, peripheral exclusion zones along the four domain walls limit the area  that is available to be occupied by  each particle in the Corner and One-wall subdomains. 
\item Observation\#2 implies that the expected value for the number of particles per subdomain should be greatest in the Center subdomain, smaller in the One-wall subdomains, and smallest in the Corner subdomains.
\end{enumerate}
While these three observations can be made about this interacting particle  system, the precise manner in which expected values of the number of particles per subdomain in the stationary response vary with both $R$ and the subdomain type is not readily ascertained, a priori. 

\subsection{Markov Chain Model}  \label{sec:MCmodel}
 We aim to identify and quantify  system properties of the directly simulated continuous model  via time series $\eta_i(t_k)\;(i=1,\ldots,9,\;t_k=k\Delta t,\;k=0,\ldots,N)$,  that track the number of particles per subdomain, where $N$ is the total number of time steps. A particle is counted as being located in the $i$th subdomain at time $t_k$ if its center is located in that subdomain at time $t_k$. Introduce a second time series  $c_i(t_k)=\eta_i(t_k)-\eta_i(t_{k-1})\;(i=1,\ldots,9, k=1,\ldots,N)$.  Assuming that $\Delta t$ is sufficiently small, this second time series will take one of three values, i.e.,
\begin{equation}
c_i(t_k) = \left\{ 
\begin{array}{ll} 
0 & \mbox{if}\;\;\eta_i(t_k) = \eta_i(t_{k-1})\\
-1 & \mbox{if}\;\;\eta_i(t_k) = \eta_i(t_{k-1})-1\\
+1 & \mbox{if}\;\;\eta_i(t_k) = \eta_i(t_{k-1})+1
\end{array}
\right.,\;\;i=1,\ldots,9,\;k=1,\ldots,N.
\label{eq:cvals}
\end{equation}
When $\Delta t$ is small, a state change in which the number of particles in a subdomain changes by more than one particle is a  rare event.  Such instances are tracked in the implementation and adverse affects can be remedied by decreasing $\Delta t$ until the resulting state changes universally obey (\ref{eq:cvals}).  

The following indicator function is also introduced to determine when the $i$th subdomain contains $j$ particles at time $t_k$:
\begin{equation}
\gamma_i^j(t_k) = \left\{ 
\begin{array}{ll} 
1 & \mbox{if}\;\;\eta_i(t_k) = j\\
0 & \mbox{if}\;\;\eta_i(t_k) \neq j
\end{array}
\right.,\;\;\;i=1,\ldots,9,\;j=0,\ldots,M,\;k=1,\ldots,N,
\label{eq:indicator}
\end{equation}
where $M\;(\leq 27)$ is the maximum number of particles present in a subdomain at any particular time. Two additional indicator functions are  then used to determine when the $i$th subdomain (containing $j$ particles at time $t_k$) gained or lost (respectively) one particle from a neighboring subdomain as time advanced from $t=t_{k-1}$ to $t=t_k$:
\begin{equation}
\beta_i^{j+}(t_k) = \left\{ 
\begin{array}{ll} 
\gamma_i^j(t_k) & \mbox{if}\;\;c_i(t_k) = +1\\
0 & \mbox{if}\;\;c_i(t_k) \neq +1
\end{array}
\right.,i=1,\ldots,9,\;j=0,\ldots,M,\;k=1,\ldots,N,
\label{eq:plusind}
\end{equation}
\begin{equation}
\beta_i^{j-}(t_k) = \left\{ 
\begin{array}{ll} 
\gamma_i^j(t_k) & \mbox{if}\;\;c_i(t_k) = -1\\
0 & \mbox{if}\;\;c_i(t_k) \neq -1
\end{array}
\right.,i=1,\ldots,9,\;j=0,\ldots,M,\;k=1,\ldots,N.
\label{eq:minusind}
\end{equation}

The quantities in (\ref{eq:indicator})-(\ref{eq:minusind}) are then used to determine probabilities for transitions between the number of particles per subdomain (states) as now outlined. Let $G_i^j$ track the number of time steps for which  the $i$th subdomain contains $j$ particles during the time interval $[t_1,t_N]$:
\begin{equation}
G_i^j=\sum_{k=1}^N \gamma_i^j(t_k),\;\;i=1,\ldots,9,\;j=0,\ldots,M.
\label{eq:Gdef}
\end{equation}
Then,  let  $\delta^{j+}_i$ and $\delta^{j-}_i$ count the number of time steps where the $i$th subdomain is in state $j$ and gains or loses (respectively) one particle:
\begin{equation}
\delta_i^{j+} = \sum_{k=1}^N \beta_i^{j+}(t_k),\;\;
\delta_i^{j-} = \sum_{k=1}^N \beta_i^{j-}(t_k),\;\;i=1,\ldots,9,\; j=0,\ldots,M.
\label{eq:pmcount}
\end{equation}
Using  (\ref{eq:Gdef})-(\ref{eq:pmcount}), it then follows that the probabilities of the $i$th subdomain gaining or losing a particle when in a particular state $j$ are, respectively:
\begin{equation}
 \gamma_i^{j+} = \frac{\delta_i^{j+}}{G_i^j},\;\;\;\;\; \gamma_i^{j-} = \frac{\delta_i^{j-}}{G_i^j},\;\;i=1,\ldots,9,\;j=0,\ldots,M.
 \end{equation}
 
These probabilities are now specialized to the subdomains depicted in Fig.~\ref{fig:subdomains}, i.e.~four corner ({\em L}) domains ($i=1,3,7,9$), four one-wall ({\em I}) domains ($i=2,4,6,8$) and one center ({\em C}) domain ($i=5$). Pooled probabilities for each of the three subdomain types are  calculated as:
 \begin{equation}
 p^{j+}_L = \frac{\sum_{i=1,3,7,9}\delta_i^{j+}}{\sum_{i=1,3,7,9}G_i^j},\;\;
  p^{j+}_I = \frac{\sum_{i=2,4,6,8}\delta_i^{j+}}{\sum_{i=2,4,6,8}G_i^j},\;\;
   p^{j+}_C = \frac{\delta_5^{j+} }{G_5^j},\;
 j=0,\ldots,M,
 \label{eq:pplus}
 \end{equation}
 for gaining a particle and as:
  \begin{equation}
 p^{j-}_L = \frac{\sum_{i=1,3,7,9}\delta_i^{j-}}{\sum_{i=1,3,7,9}G_i^j},\;\;
  p^{j-}_I = \frac{\sum_{i=2,4,6,8}\delta_i^{j-}}{\sum_{i=2,4,6,8}G_i^j},\;\;
   p^{j-}_C = \frac{\delta_5^{j-} }{G_5^j},\;
 j=0,\ldots,M,
 \label{eq:pminus}
 \end{equation}
 for losing a particle. 

\begin{figure}[h] 
\begin{center}
\includegraphics[scale=0.40]{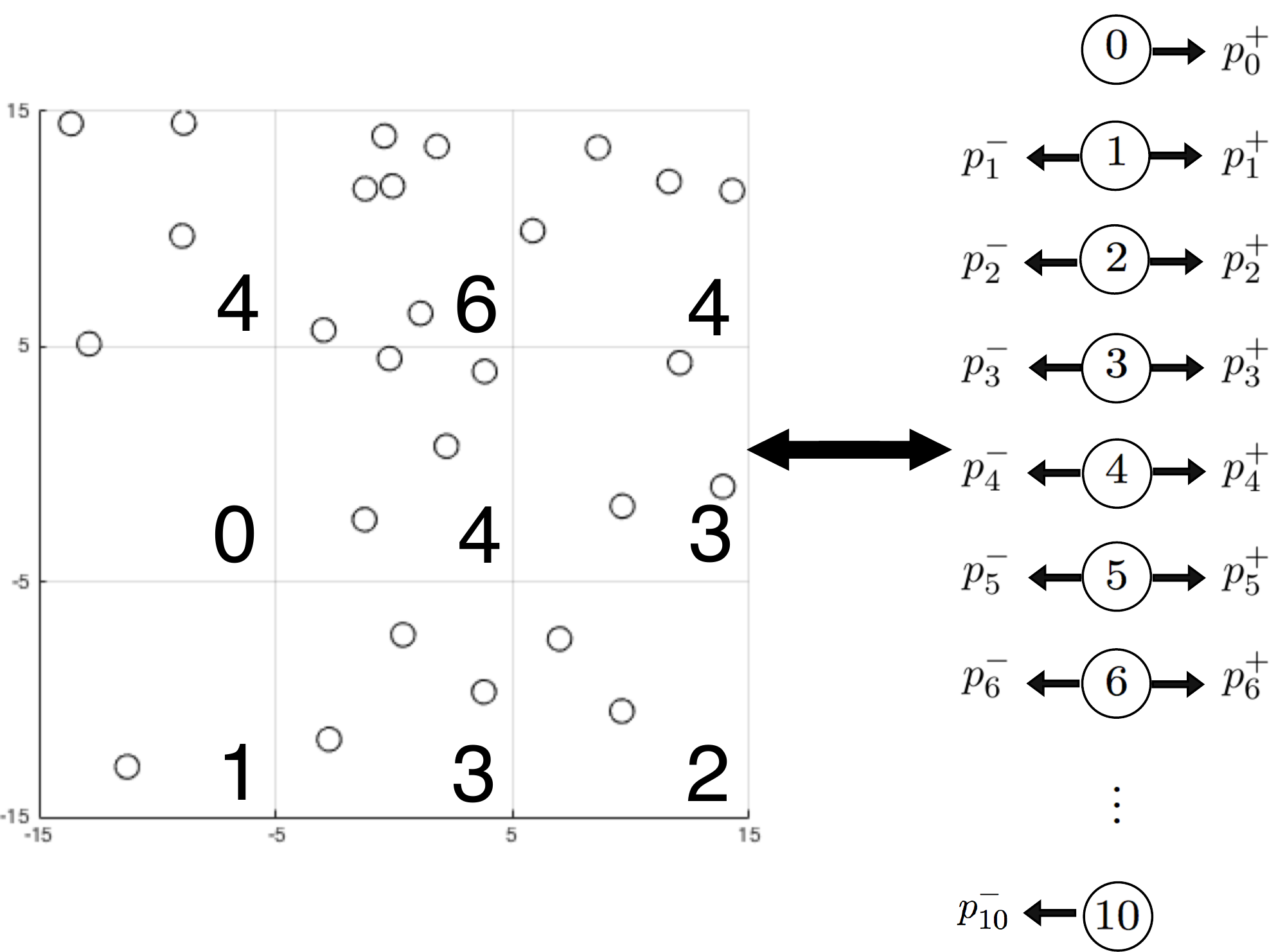}
\end{center}
\vspace{-.1truein}
\caption{Illustration of the Markov chain model for tracking transition probabilities in the number of particles per subdomain at a snapshot in time when $N_s=10$. Note that transition probabilities are calculated separately for each of the three subdomain types $\alpha=L,I,C$ using equation (\ref{eq:psame}).}
\label{fig:grid2prob}
\end{figure}

By pooling data from (\ref{eq:pplus})-(\ref{eq:pminus}) for each of the three subdomain types, the probability that a subdomain of type $\alpha$ remains in state $j$ in the Markov chain model is:
\begin{equation}
p^{j*}_\alpha = 1 - p^{j+}_\alpha - p^{j-}_\alpha,\;\;\alpha=L,I,C,\;\;j=0,\ldots,N_s,
\label{eq:psame}
\end{equation}
where $N_s(\leq M)$ is the maximum number of particles per subdomain to be accounted for via the states in the Markov chain model.   As $N_s$ increases past $8$-$9$ particles per subdomain, the occurrence of such states in a subdomain becomes quite rare.  Hence, the model is calibrated to choose a value for $N_s$ beyond which the associated results exhibit little sensitivity to further increasing $N_s$. For illustration, the process of tracking state transitions in the case $N_s=10$ is shown in Fig.~\ref{fig:grid2prob}.

The probabilities in equations (\ref{eq:pplus})-(\ref{eq:psame}) are used to assemble three tridiagonal transition matrices for the Markov Chain model: 
 \begin{equation}
 P_\alpha = \begin{bmatrix}
    p^{0*}_\alpha & p^{0+}_\alpha & 0 & \dots  & 0 \\
   p^{1-}_\alpha & p^{1*}_\alpha & p^{1+}_\alpha & \dots  & 0 \\
    \vdots & \ddots & \ddots & \ddots & \vdots \\
    0  & \dots & p^{(N_s-1)-}_\alpha   & p^{(N_s-1)*}_\alpha  & p^{(N_s-1)+}_\alpha \\
   0  & \dots  & 0 & p^{N_s-}_\alpha  & p^{N_s*}_\alpha
\end{bmatrix},\;\;\alpha=L,I,C.
\label{eq:transMat}
\end{equation}
Lastly, stationary distributions of the three types of transition probability vectors ($\vec{\pi}_C$, $\vec{\pi}_I$, $\vec{\pi}_L$) are computed as the solutions to the linear algebraic equations:
\begin{equation}
	\vec{\pi}_C = \vec{\pi}_C P_C,\;\;\;\;\;\;
	\vec{\pi}_I = \vec{\pi}_I P_I,\;\;\;\;\;\;
	\vec{\pi}_L = \vec{\pi}_L P_L.
\label{eq:statdist}
\end{equation}
The solutions of (\ref{eq:statdist}) then, effectively, constitute a surrogate Markov chain model for the directly simulated continuous model.  These solutions also facilitate a detailed investigation of both the expected values and uncertainty for the number of particles per subdomain in the stationary regime as both the subdomain type and the particle radius are varied.

\section{Results}
\label{sec:main}
We first illustrate some key properties of the continuous model used to directly simulate particle interactions and discuss its calibration (Sec.~\ref{sec:contRes}).  Results are then compared to those obtained using the  surrogate Markov chain model, and also used to study the stationary response for the number of particles per subdomain, quantifying both expected values and uncertainty with increasing particle radius (Sec.~\ref{sec:MCRes}).

\subsection{Continuous Model Results}
\label{sec:contRes}
To quantify uncertainty in the continuous model,  simulations were run over a large number of realizations of the initial particle configuration on a square of side length $30$. This length scale was normalized relative to a particle radius $R$, i.e.~with $0<R\leq 1$. 
A large number of time steps was also required to be in a regime where the average number of particles per subdomain exhibited a stationary response. 
This process was then repeated as the particle radius $R$ was varied in the range $[0.1,0.9]$ to determine effects of particle size on the quantity of interest. 
For each realization, each of the nine subdomains initially contained  three particles.  These initial particle positions were not varied across realizations.   Particles were prescribed  random initial velocities that were fixed in the sense that they were all prescribed the same initial speed $|{\bf v}|=8$, but their  initial directions were drawn from a uniform distribution.  At each time step, the positions of all 27 particles were updated and the center of each particle was used to determine if the subdomain in which the particle resided at the prior time step had changed.  Conservation of linear momentum was used to determine particle locations when a collision with another particle or with one of the four rigid walls occurred over the duration of one time step. The resulting set of particle locations  was then used to compute the time series $\eta_i(t_k)$ in Sec.~2.2 that track the number of particles per subdomain as time advances. 

\begin{figure}
\begin{center}
\includegraphics[scale=0.3]{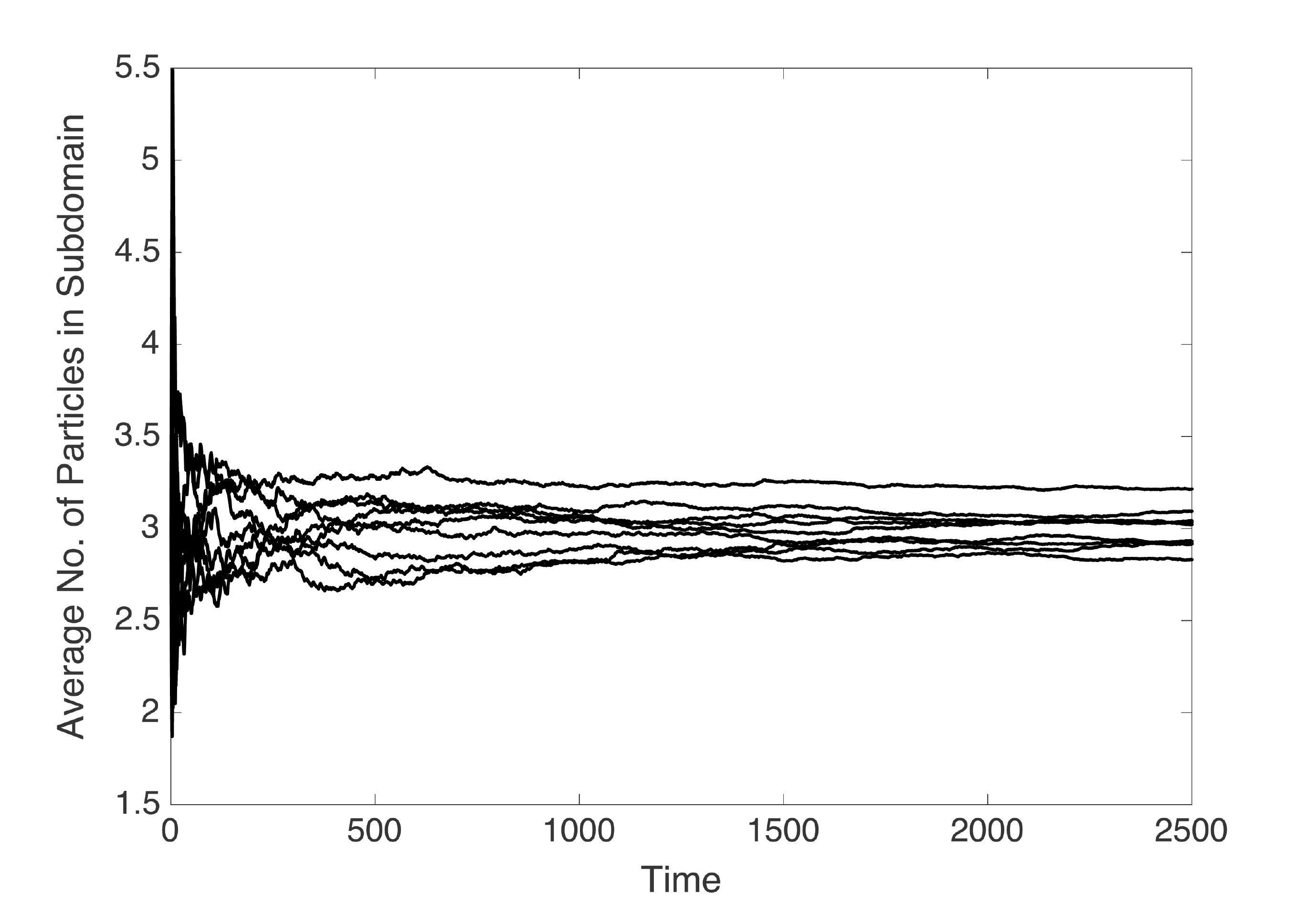}
\end{center}
\caption{Illustration of the stationary response for the average number of particles per subdomain in the case of 27 particles over the 9 subdomains shown in Fig.~\ref{fig:subdomains}.  Results are shown for a single realization in the case $R=0.5$ and $\Delta t = 0.0125\;(N=2\times 10^5)$. }
\label{fig:continAvgBis}
\end{figure}

\begin{figure} 
	\begin{center}
	   \includegraphics[scale=0.4]{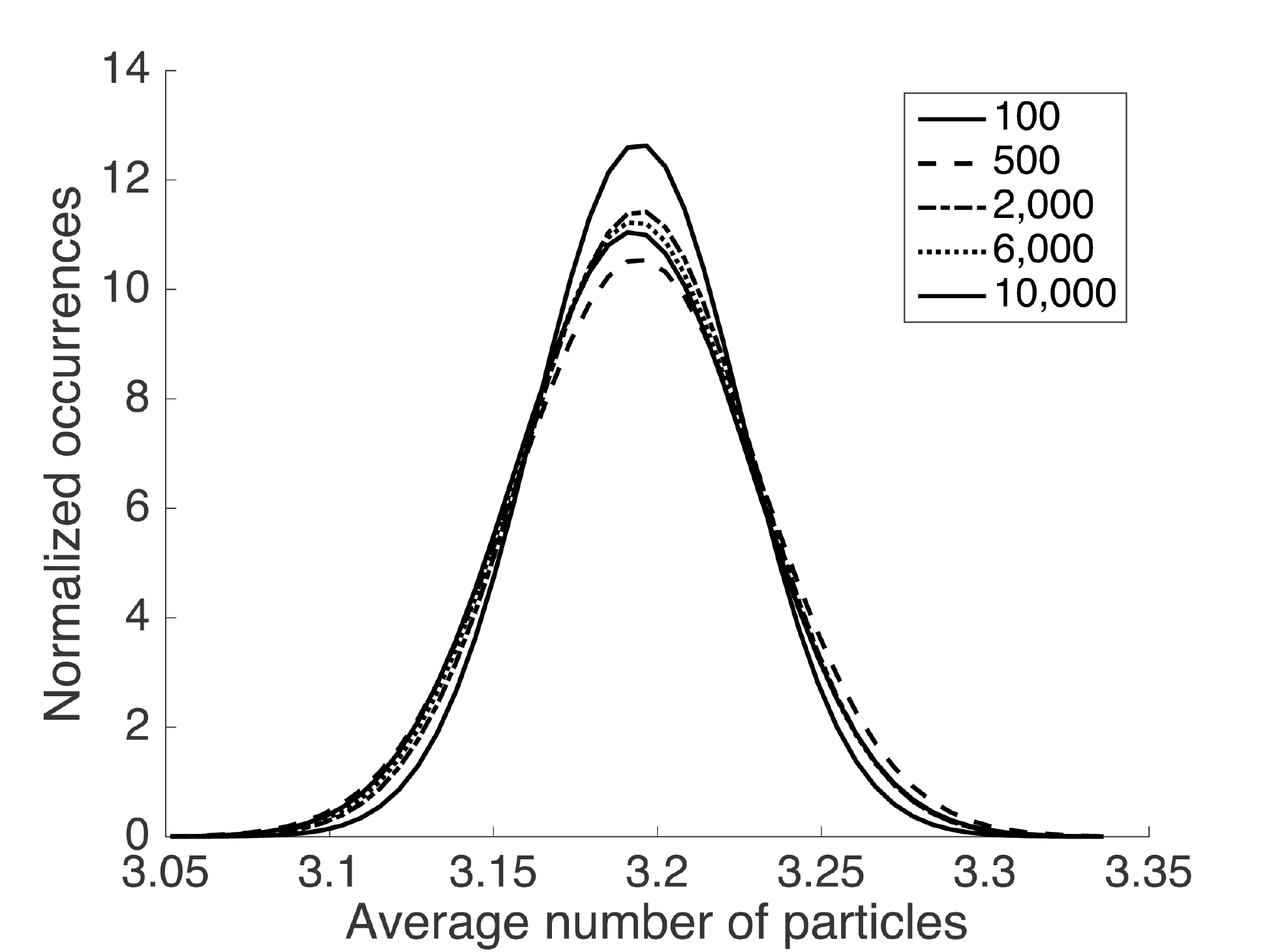}
	\end{center}
	\caption{Illustration of the effect on histograms for the number of particles per subdomain $\bar{\eta}_C$ as the number of realizations is increased.  Results are shown for the center subdomain in the case $R=0.5$.}
	\label{fig:cencontinOVL}
\end{figure}

It was  determined that a value of $N=2\times 10^5$ (time steps), corresponding to $\Delta t = 0.0125$,  was sufficient for yielding data exhibiting the stationary properties needed to build the surrogate Markov chain model.  This value of $N$ was used in all subsequent simulations.  The stationary nature of the average number of particles in each of the nine subdomains is illustrated for a single realization in Fig.~\ref{fig:continAvgBis} in the case $R=0.5$.  Mean values of  $\bar{\eta}_\alpha\;(\alpha=L,I,C)$ of $\eta_i(t_k)\;(t_k=k\Delta t, k=1,\ldots,N)$ were obtained by averaging over all times  and then pooling data for each of the three subdomain types, i.e.~corner ($\alpha=L\leftrightarrow i=1,3,7,9$), one-wall ($\alpha=I\leftrightarrow i=2,4,6,8$) and center ($\alpha=C\leftrightarrow i=5$) across all realizations.   As illustrated in Fig.~\ref{fig:cencontinOVL}, 6,000 realizations were sufficient to stabilize variation in this statistic for the quantity of interest.  Specifically, in this regime the mean and standard deviation in the normal distribution stabilized their values to 3.19 and 0.036, respectively, for the case shown in Fig.~\ref{fig:continAvgBis}. Consequently, 6,000 realizations were used in all subsequent simulations. 

\begin{figure} 
	\begin{center}
	   \includegraphics[scale=0.65]{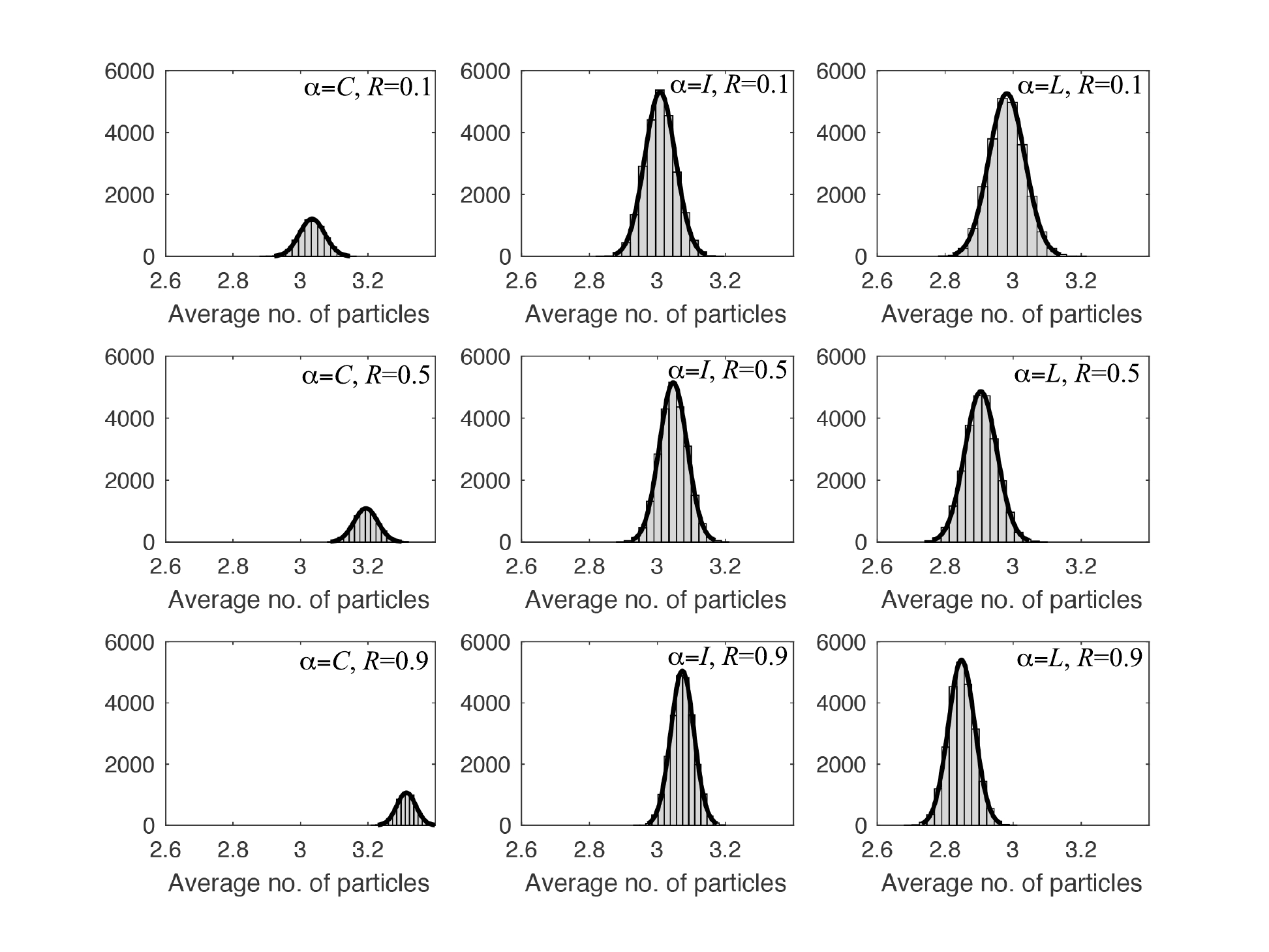}
	\end{center}
	\caption{Histograms and fitted normal distributions for mean values of the number of particles per subdomain $\bar{\eta}_\alpha$ with varying subdomain type ($\alpha=L,I,C$) and particle radius $R$ obtained by direct simulation using the continuous model.}
	\label{fig:continHist}
\end{figure}

\begin{figure}
\begin{center}
	\includegraphics[scale=0.23]{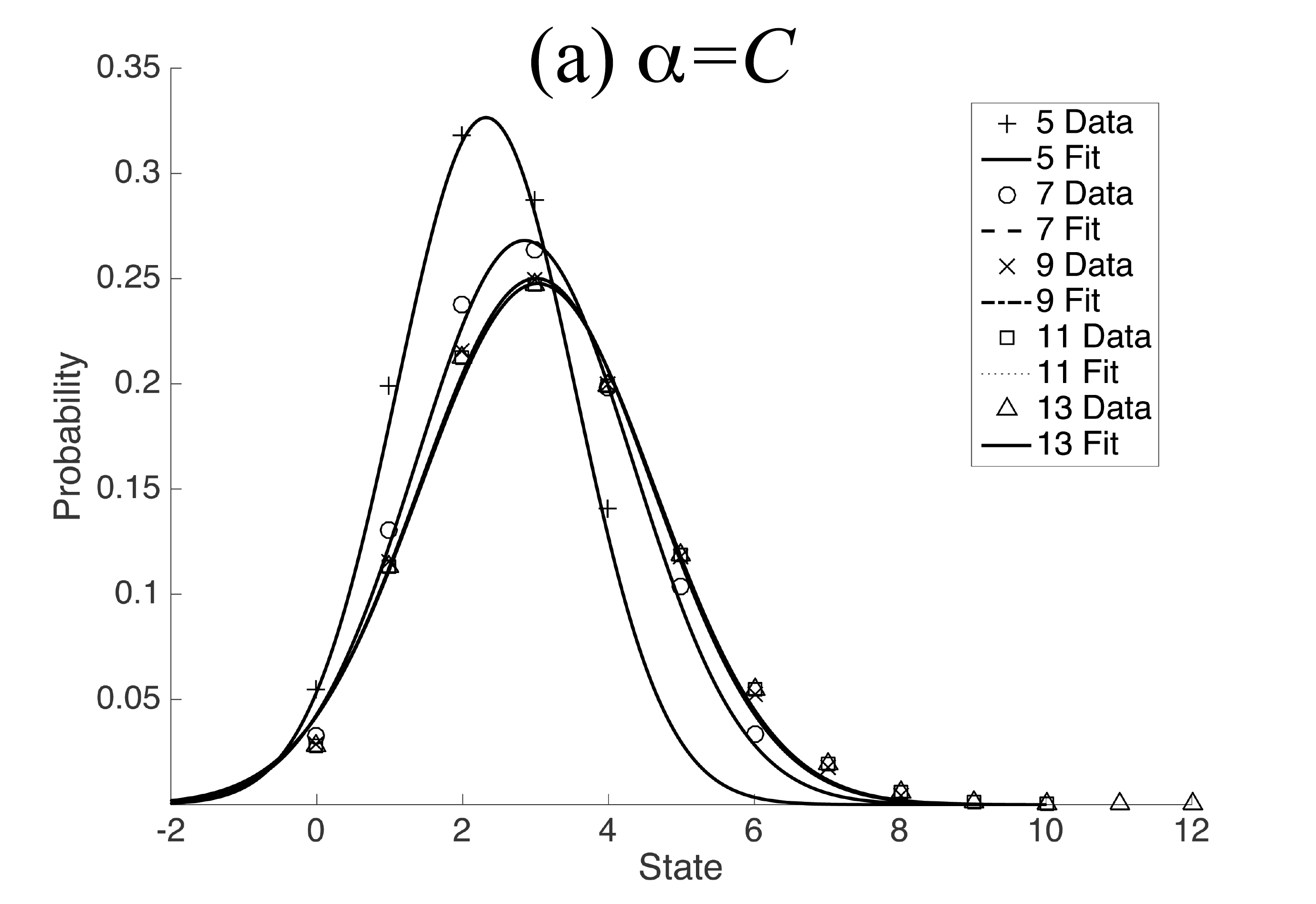}
	\includegraphics[scale=0.23]{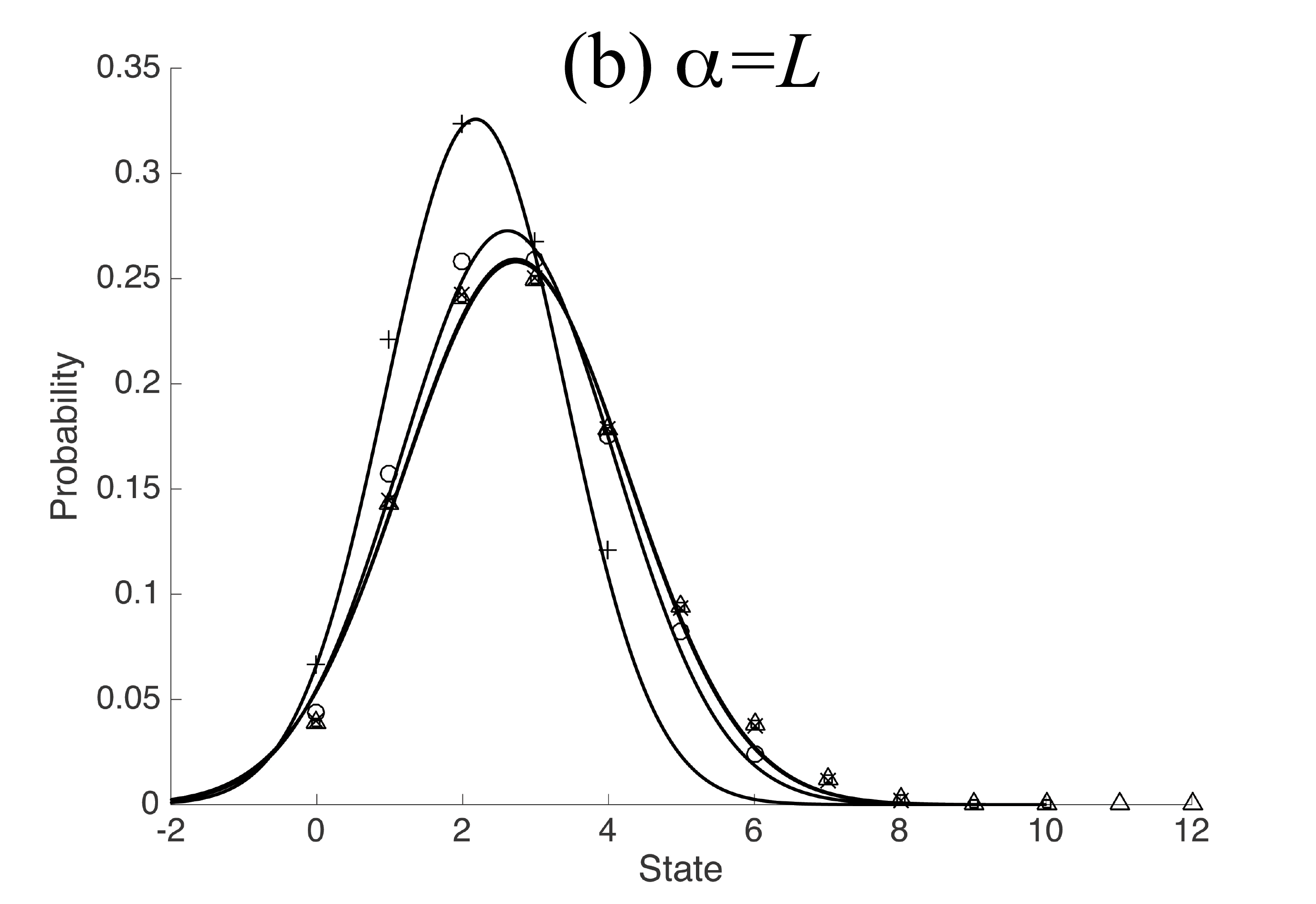}
	\includegraphics[scale=0.23]{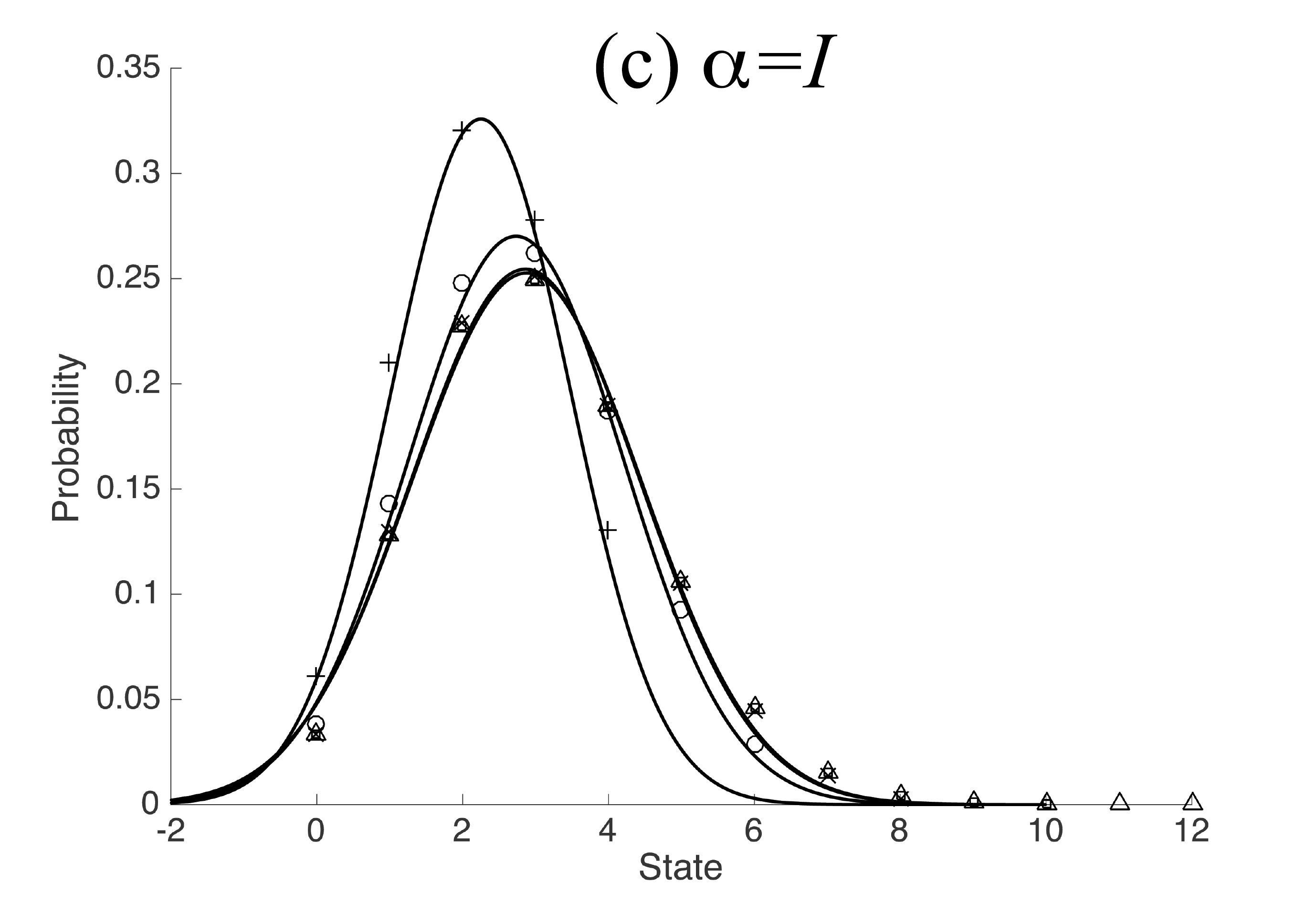}
\end{center}
	\caption{Effects of varying the maximum number of states per subdomain $N_s$ in the Markov chain model.  Stationary distributions obtained from (\ref{eq:statdist}) were fit with the probability density function (\ref{eq:truncDist}) subject to the constraint (\ref{eq:constraint}) in the case $R=0.5$: (a) Center ($\alpha=C$), (b) Corner ($\alpha=L$), (c) One-wall ($\alpha=I$).}
	\label{fig:TN_states}
\end{figure}

\begin{figure}
\begin{center}
		\includegraphics[scale=0.18]{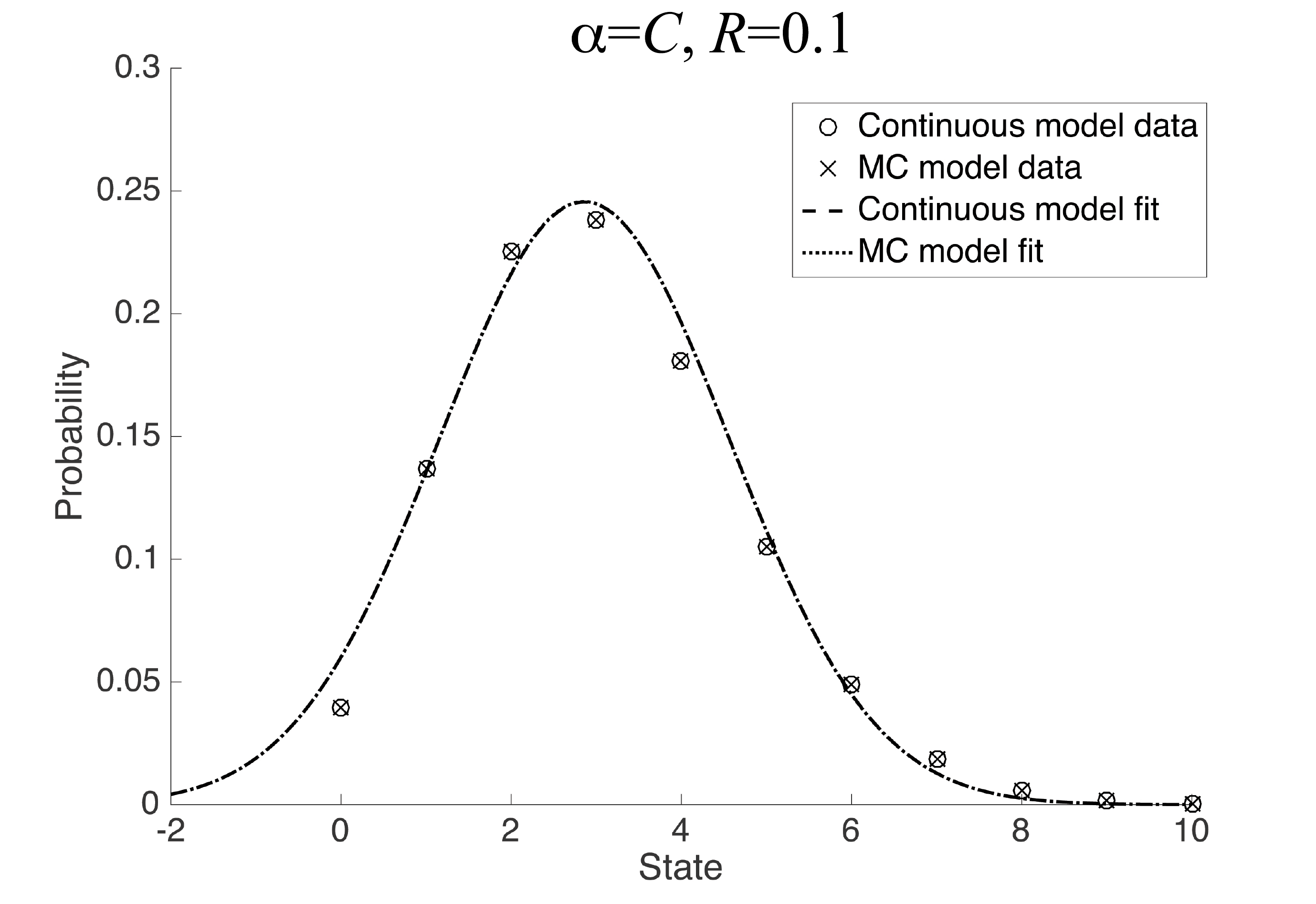}
		\includegraphics[scale=0.18]{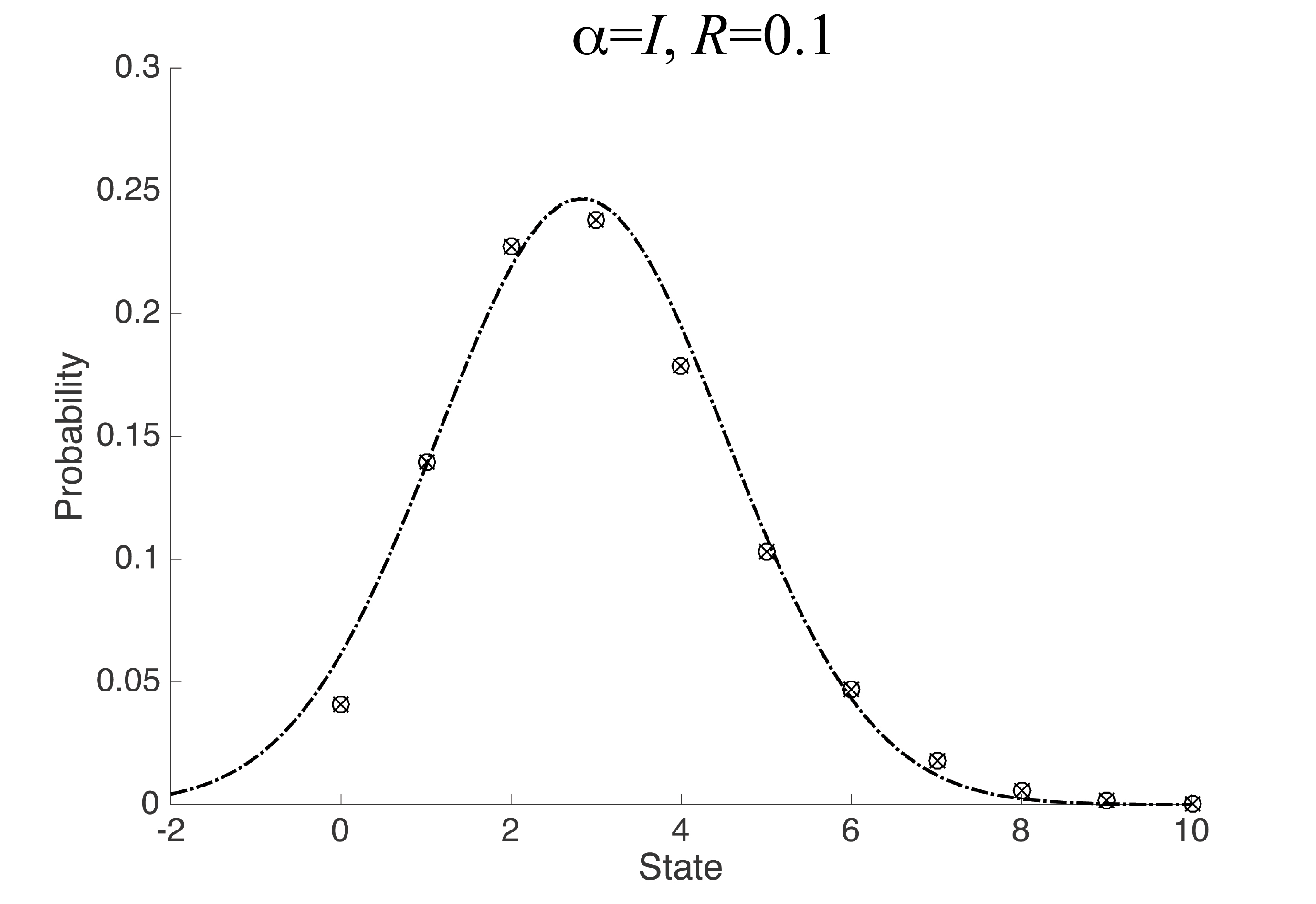}
		\includegraphics[scale=0.18]{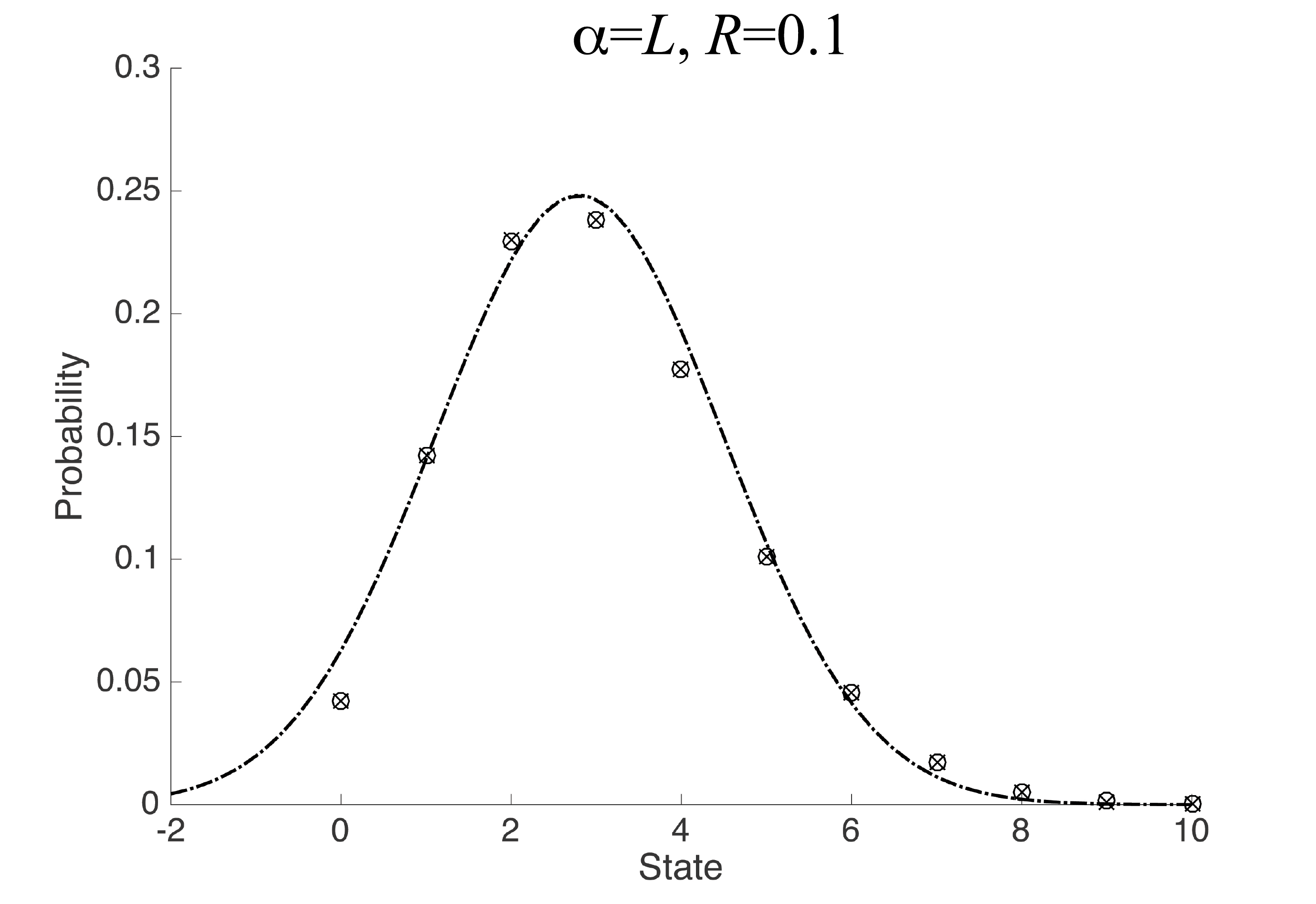}
	
		\includegraphics[scale=0.18]{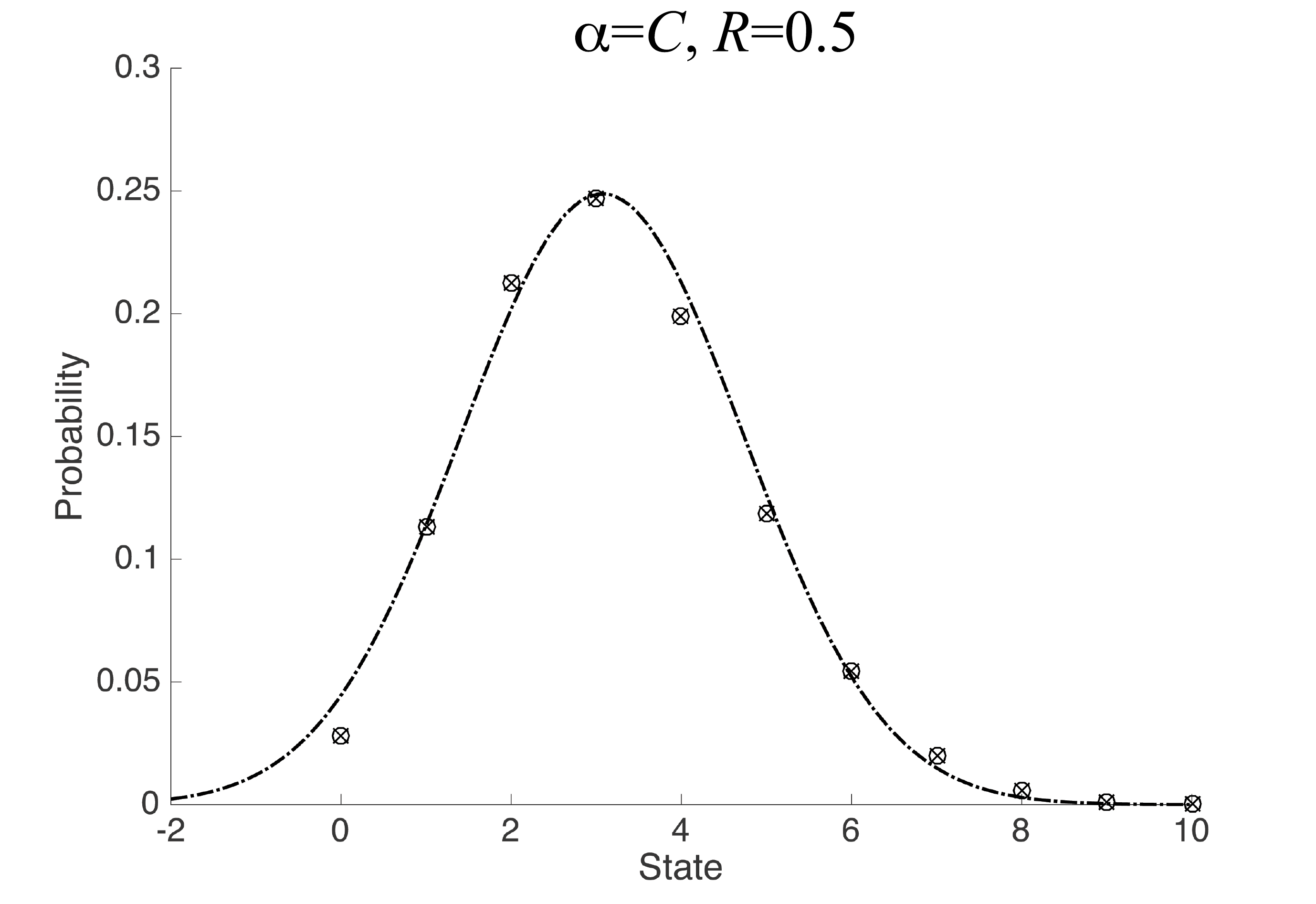}
		\includegraphics[scale=0.18]{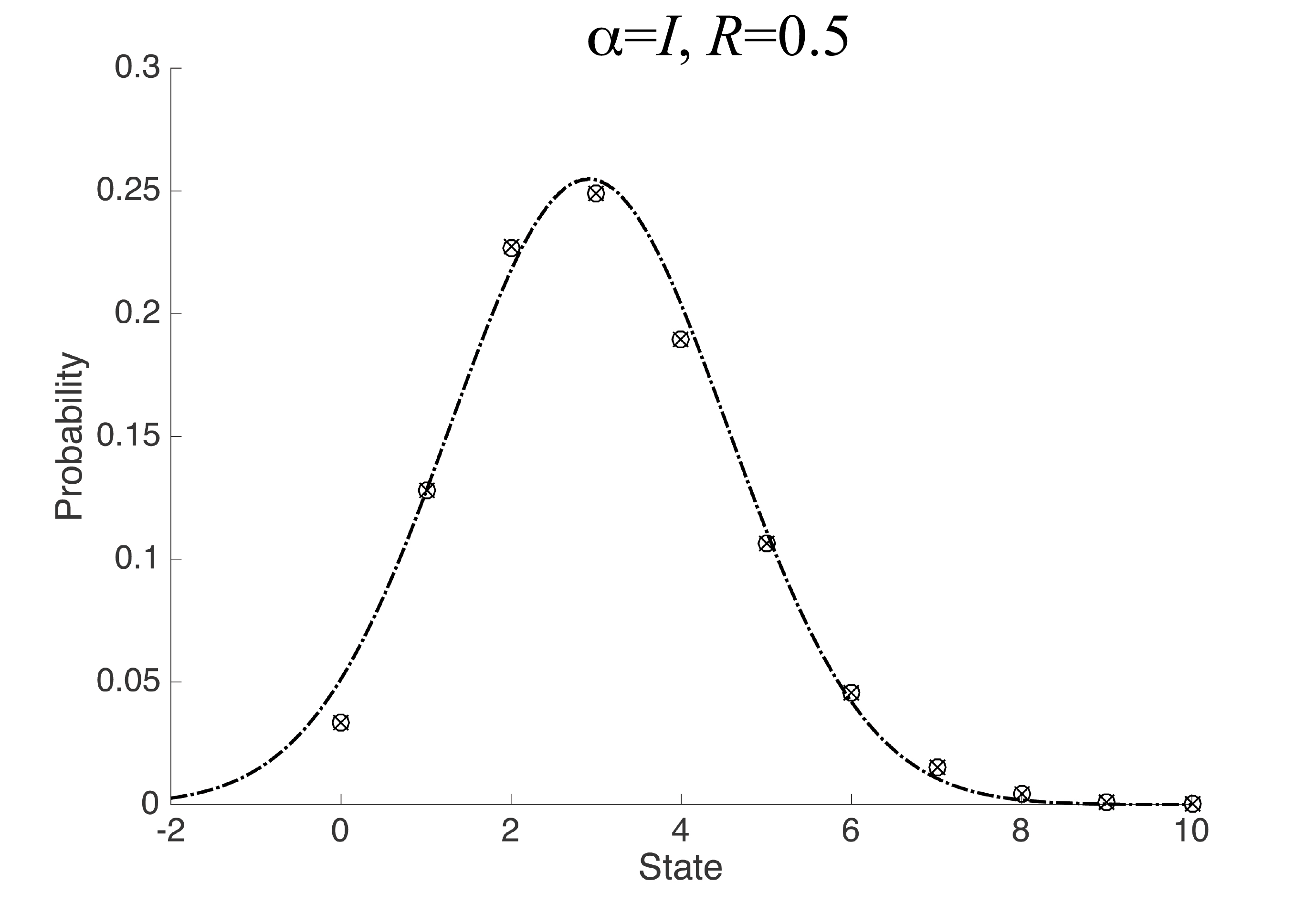}
		\includegraphics[scale=0.18]{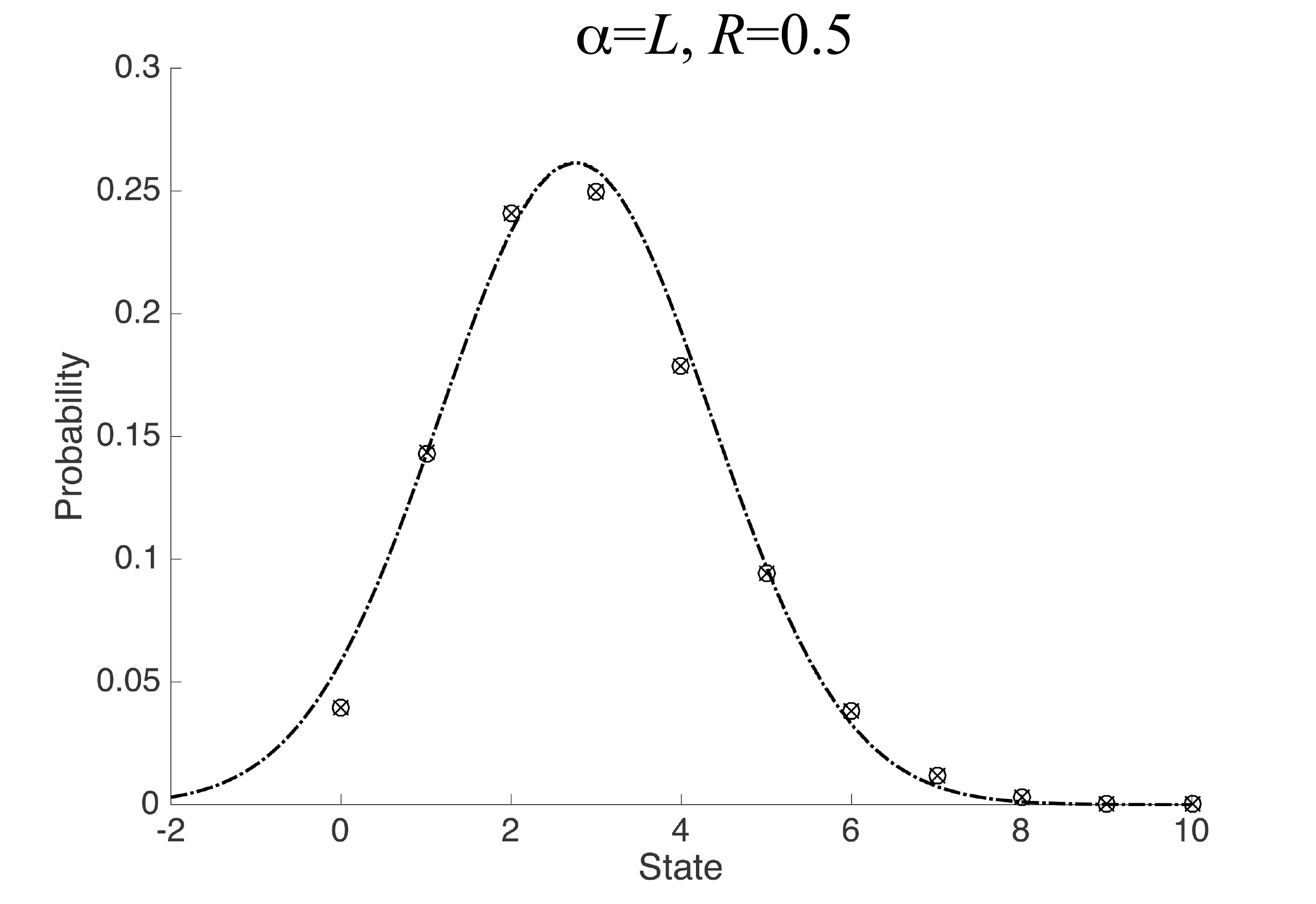}
		
		\includegraphics[scale=0.18]{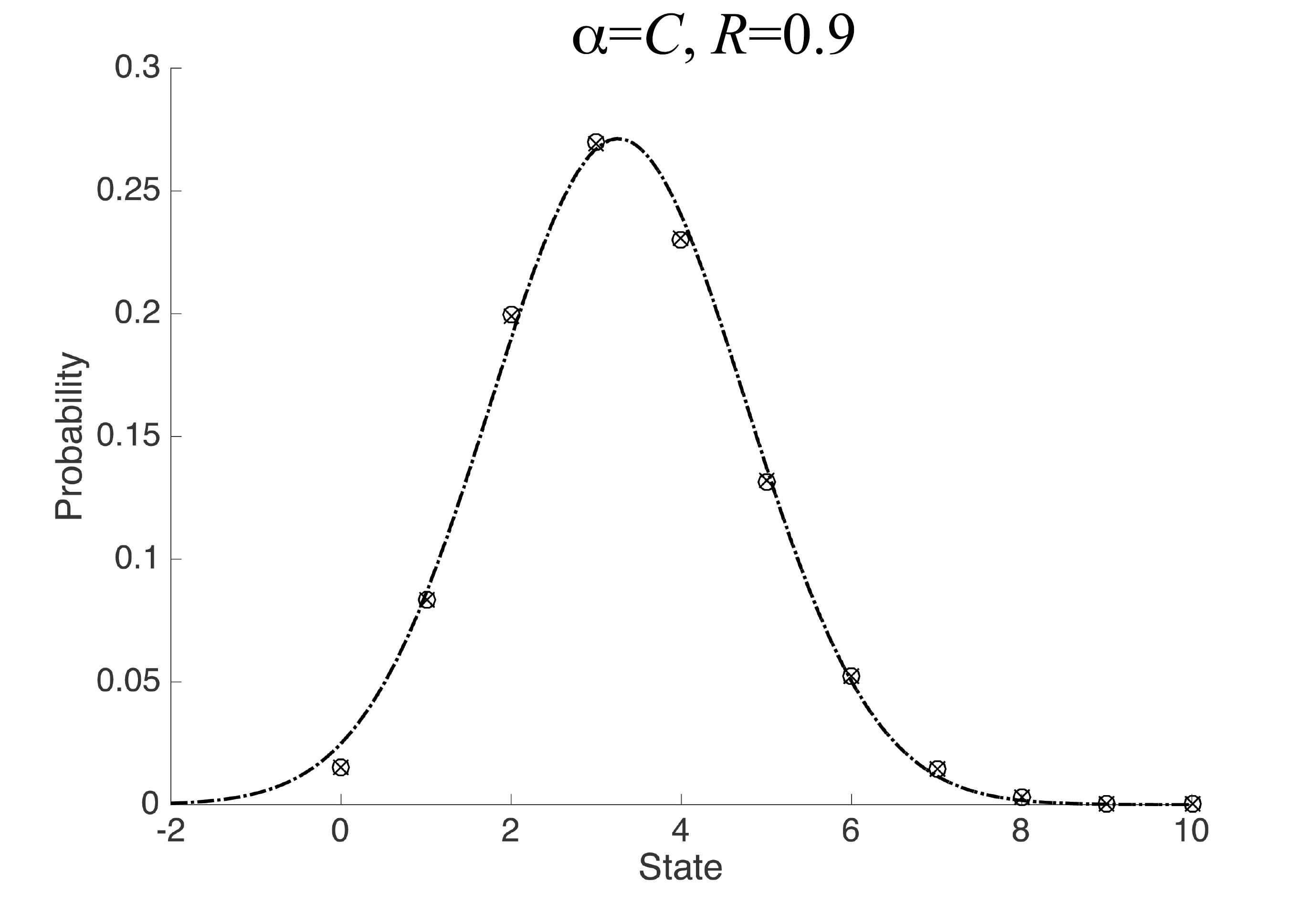}
		\includegraphics[scale=0.18]{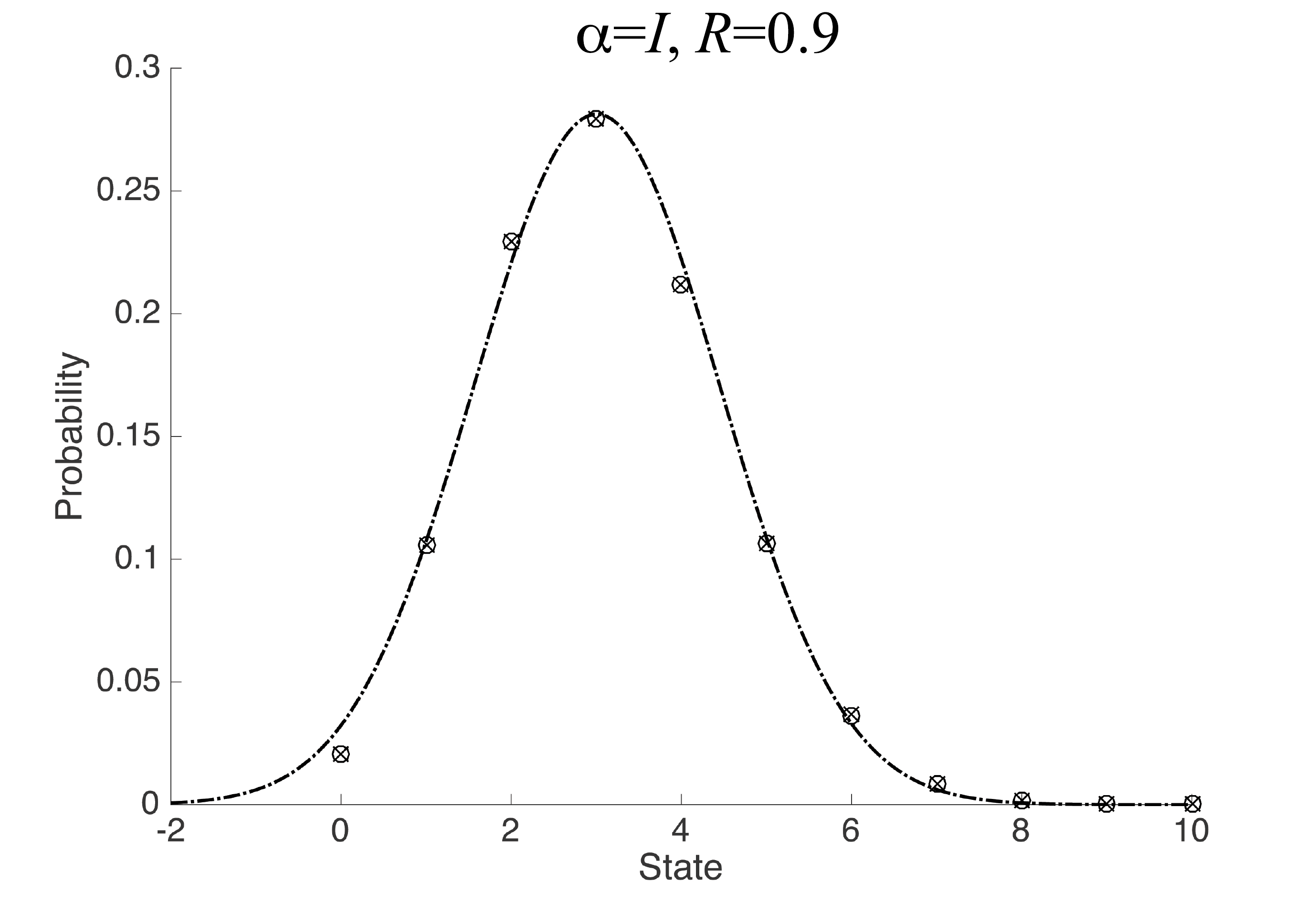}
		\includegraphics[scale=0.18]{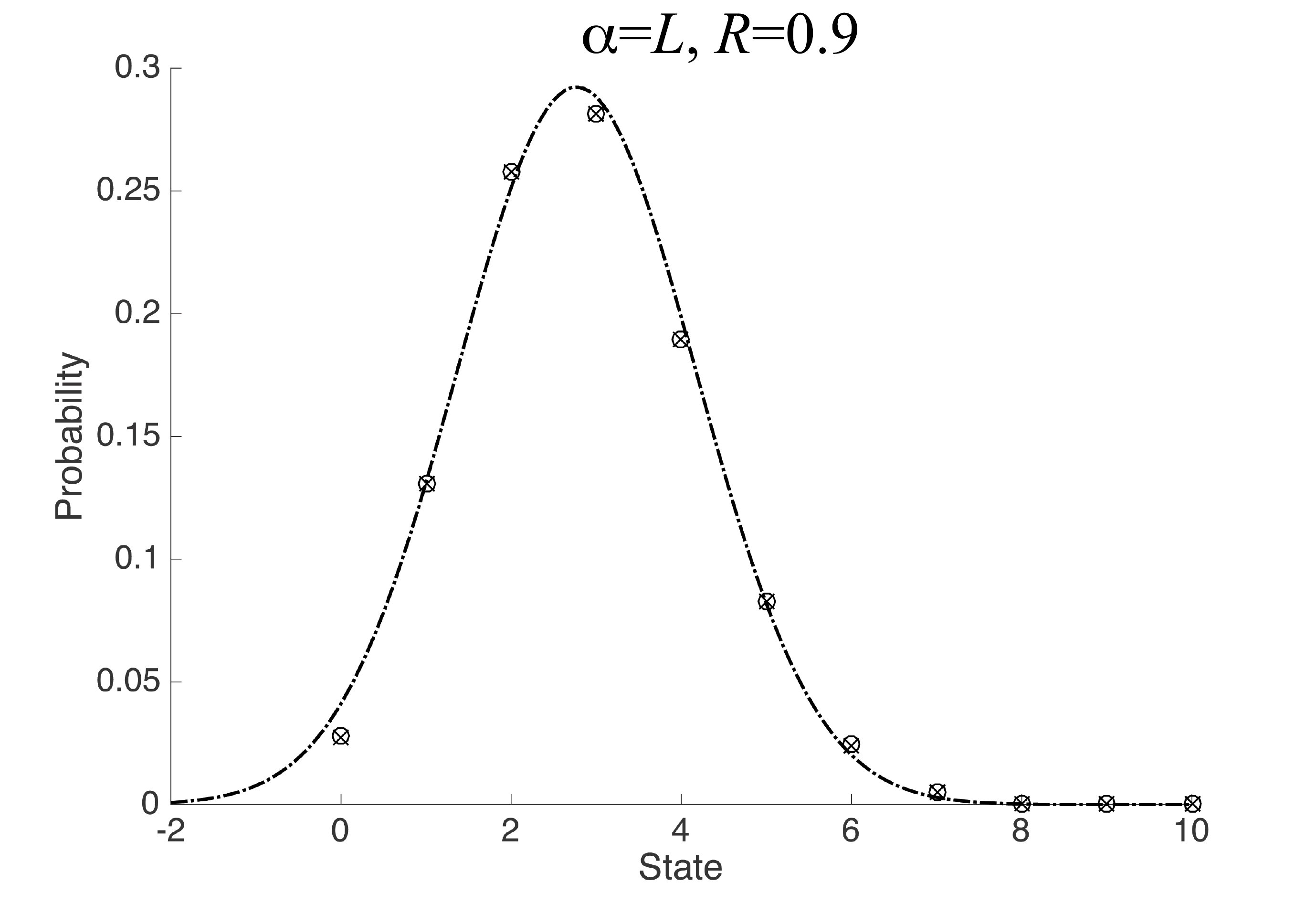}
\end{center}
		\caption{Distributions for the number of particles in each subdomain type ($\alpha=L,I,C$) with varying particle radius $R$.  Stationary distributions $\vec{\pi}_\alpha$  from  the Markov Chain (surrogate) model are compared to  data obtained using the (continuous) direct simulation model. The surrogate model was fit using the truncated normal distribution (\ref{eq:truncDist}) subject to the constraint (\ref{eq:constraint}).}
		\label{fig:TNall}
\end{figure}

Via direct simulation using the continuous model, histograms for the average number of particles per subdomain $\bar{\eta}_\alpha$ were determined as the particle radius $R$ was varied between $0.1$ and $0.9$.  Results pooled for each of the three subdomain types indicated that this statistic appeared to follow a normal distribution (Fig.~\ref{fig:continHist}).  These results also illustrate observation 3 from Sec.~2.1, i.e.~that the Center ($\alpha=C$) subdomain (left column) has, on average, a greater number of particles than the One-wall ($\alpha=I$) subdomains  (middle column) which, in turn, have a greater number of particles that the Corner ($\alpha=L$) subdomains (right column). It is also observed (Fig.~\ref{fig:continHist}) that these differences become more pronounced as the particle radius $R$ was increased from a value of 0.1 to a value of 0.9. 

Overall, these results for the continuous model provide data, obtained via direct  simulation into the stationary regime  over many realizations, for formulation of the Markov chain (surrogate) model described in Sec.~2.2.  They also illustrate fundamental statistical properties for the quantity of interest with respect to particle radius $R$ and subdomain type $\alpha=L,I,C$  that will also be used to evaluate accuracy of the surrogate model. 

\subsection{Markov Chain Model Results} \label{sec:MCRes}
Based on direct numerical simulations performed using the continuous model, the Markov Chain model was formulated by calculating the transition probabilities in (\ref{eq:pplus})-(\ref{eq:transMat}).  The matrices in (\ref{eq:transMat}) were then used to determine the three stationary distributions $\vec{\pi}_\alpha,\alpha=L,I,C$ in (\ref{eq:statdist}).  A qualitative analysis of the results indicated that a {\em truncated normal distribution} (Johnson et al.~\cite{Johnson}) was well-suited for fitting a probability density function to all three stationary distributions.  This distribution is represented as:
\begin{equation}
	f(x;\mu,\sigma,a,b) = \frac{\phi(\frac{x-\mu}{\sigma})}{\sigma \left( \Phi \left(\frac{b - \mu}{\sigma}\right) - \Phi \left( \frac{b - \mu}{\sigma} \right) \right) },\;\;a \leq x \leq b \leq \infty,
	\label{eq:truncDist}
\end{equation}
where:
\begin{equation}
\phi(\xi) =\frac{1}{\sqrt{2 \pi}} \exp \left( -\frac{1}{2} \xi^2 \right)\mbox{  and  }\Phi(x) = \frac{1}{2} (1 + \text{erf} ( x/ \sqrt{2})).
\label{eq:PDFs}
\end{equation}
In (\ref{eq:PDFs}), $\phi(\xi)$ and $\Phi(x)$ are the probability density function and the cumulative distribution function of a standard normal distribution, respectively and $[a,b]$ is the range of the random variable being considered.

\begin{figure}
\begin{center} 
		\includegraphics[scale=0.18]{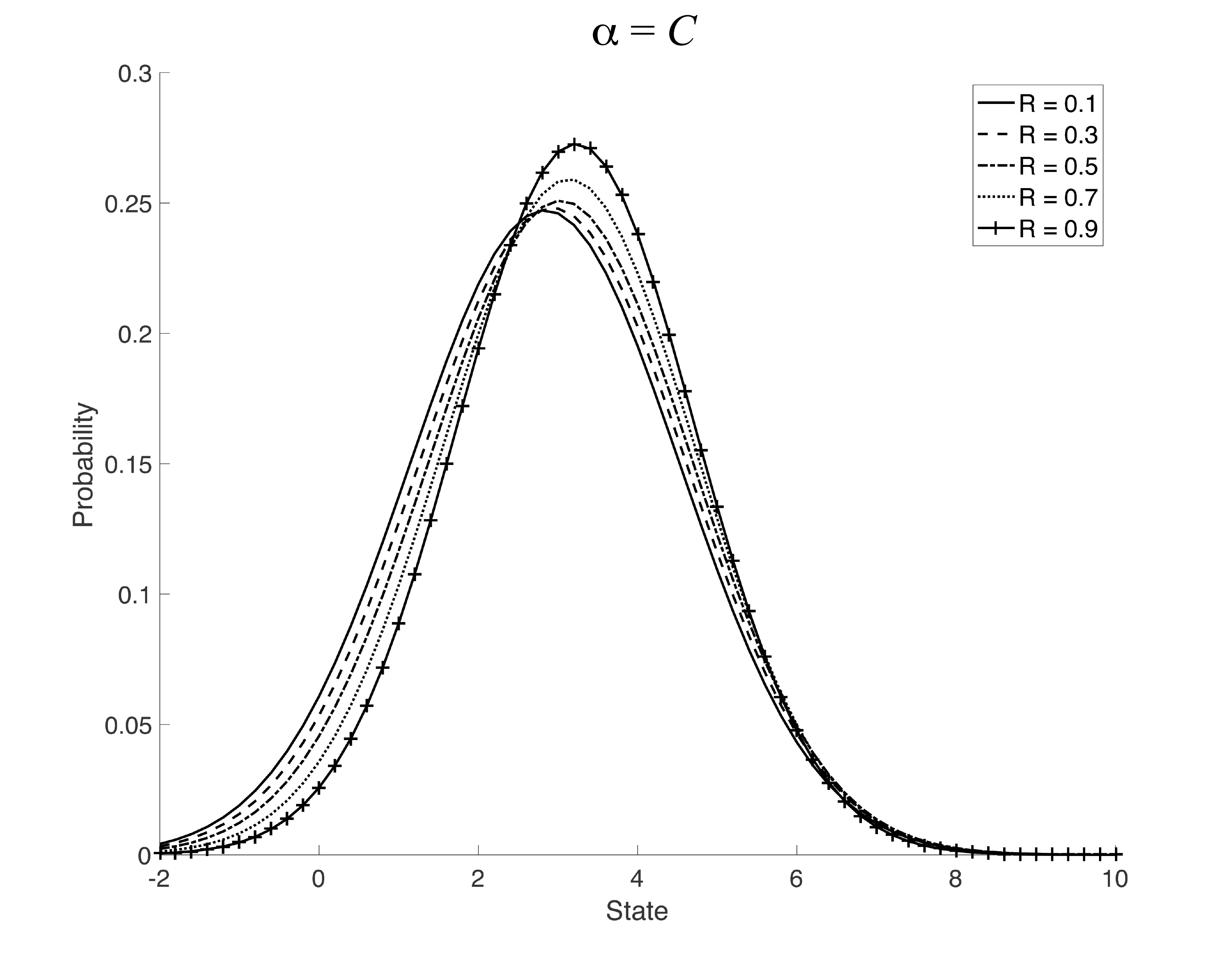}
		\includegraphics[scale=0.18]{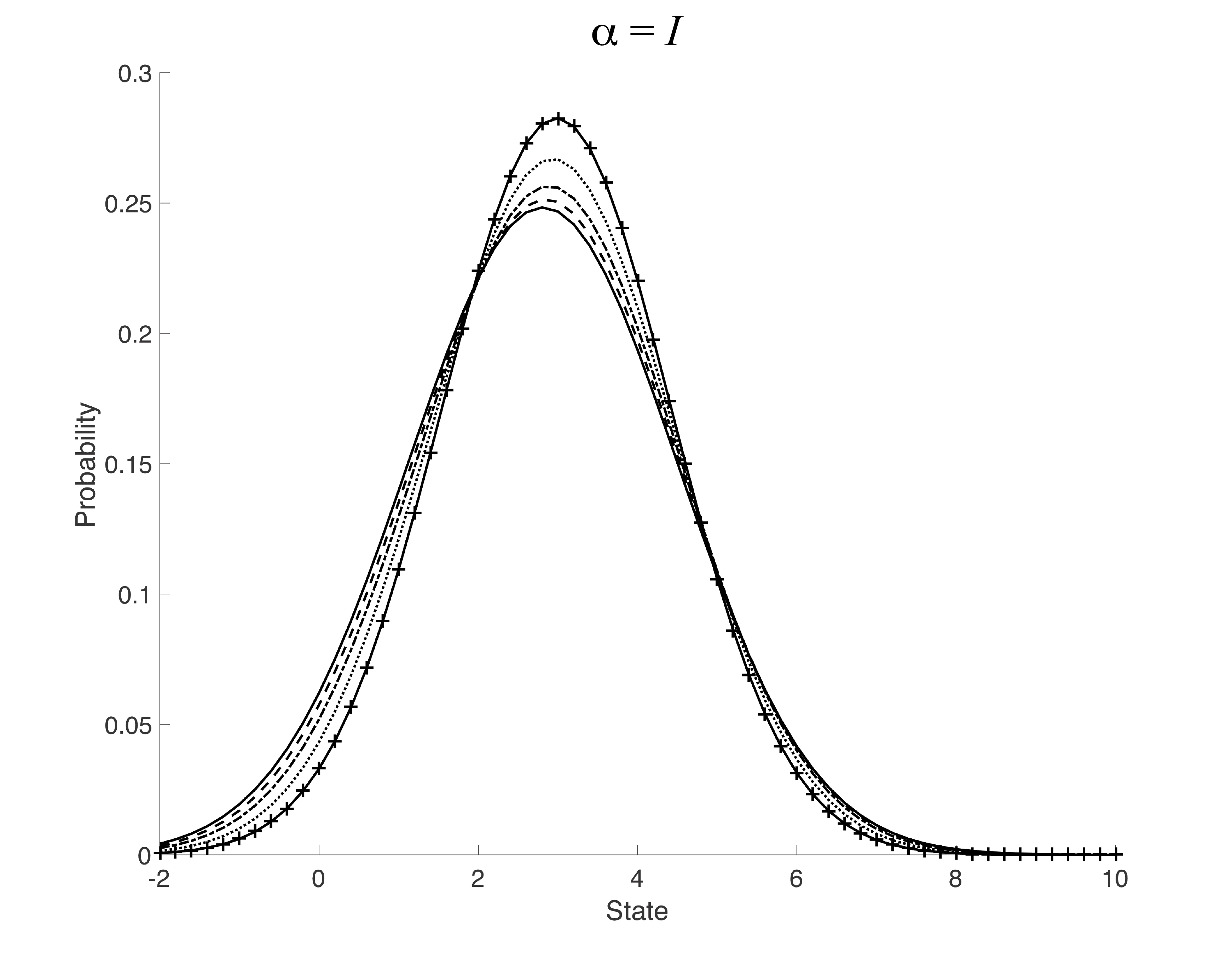}
		\includegraphics[scale=0.18]{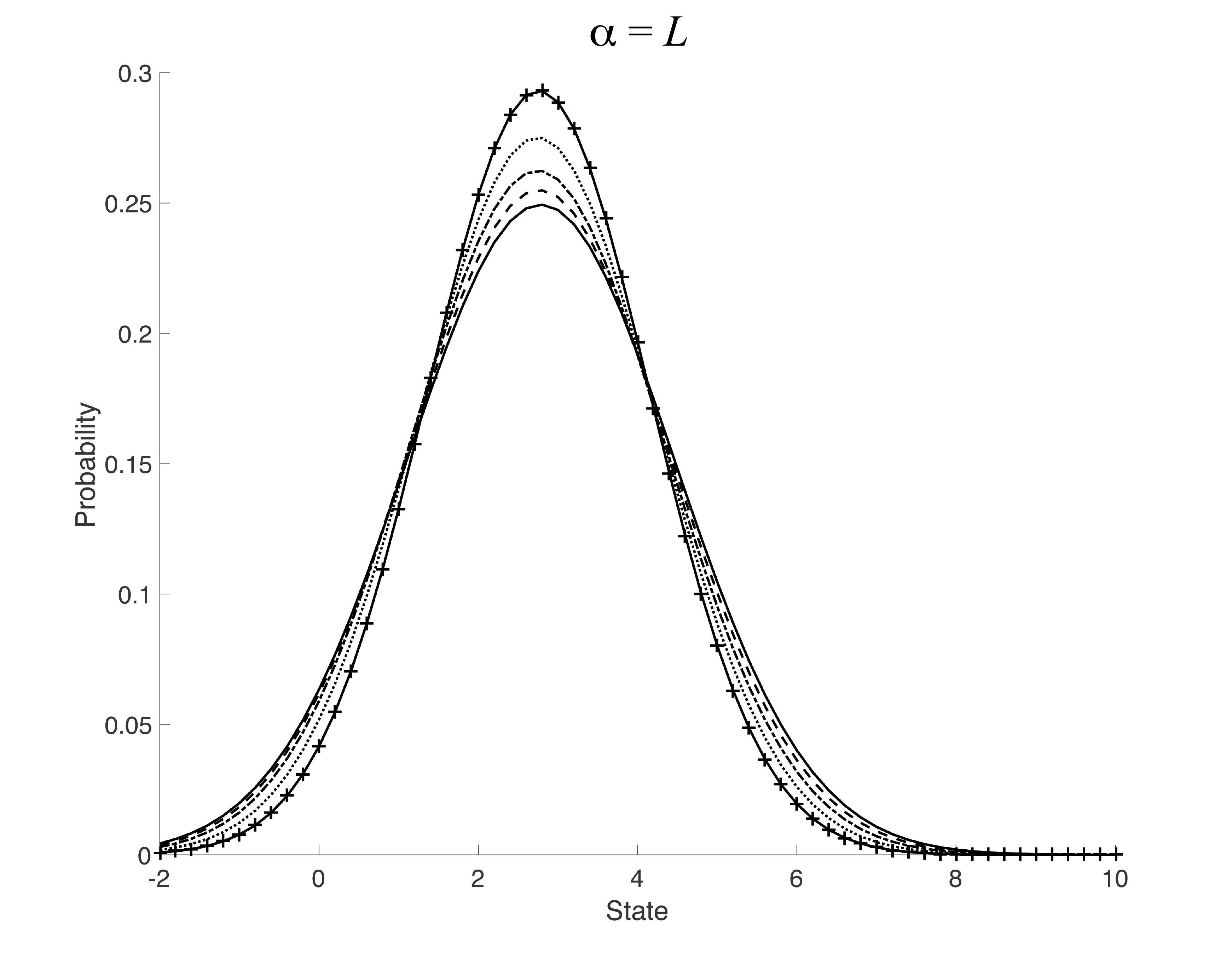}
\end{center}
\caption{Distributions for the number of particles in each subdomain grouped by subdomain type ($\alpha=L,I,C$) to illustrate effects of varying particle radius $R$.  Stationary distributions $\vec{\pi}_\alpha$  from  the Markov Chain (surrogate) model were fit using the truncated normal distribution (\ref{eq:truncDist}) subject to the constraint (\ref{eq:constraint}).}
\label{fig:TNR}
\end{figure}

\begin{table}[]
	\centering
	\begin{tabular}{|l|ll|ll|ll|}
		\hline
		\multicolumn{7}{|l|} {Continuous Model}\\
		\hline
		\multicolumn{1}{|c|}{Radius} & \multicolumn{2}{c|}{Center}                               & \multicolumn{2}{c|}{One-wall}                             & \multicolumn{2}{c|}{Corner}                               \\ \hline
		\multicolumn{1}{|c|}{}       & \multicolumn{1}{c}{$\mu$} & \multicolumn{1}{c|}{$\sigma$} & \multicolumn{1}{c}{$\mu$} & \multicolumn{1}{c|}{$\sigma$} & \multicolumn{1}{c}{$\mu$} & \multicolumn{1}{c|}{$\sigma$} \\ \hline
		0.1                          & 2.8578		    &1.7037		    &2.8287 		   &1.6967			    &2.8002   		 &1.6904 \\
    		0.3			     &	2.9632		    &1.6869		    &2.8681		   &1.6670			    &2.7754		  &1.6473 \\
   		0.5		             & 3.0708		    & 1.6559		    &2.9104		   &1.6247			    &2.7545    		&1.5924 \\
    		0.7		  	    & 3.1758		    & 1.5869		   & 2.9603		   & 1.5463		    &2.7554		   &1.5057\\
		0.9			    & 3.2613    		    &1.4926		   & 3.0074		   &1.4453 		     &2.7682		   & 1.3981                     \\ \hline
		\multicolumn{7}{|l|} {Markov Chain model}\\
		\hline
		\multicolumn{1}{|c|}{Radius} & \multicolumn{2}{c|}{Center}                               & \multicolumn{2}{c|}{One-wall}                             & \multicolumn{2}{c|}{Corner}                               \\ \hline
		\multicolumn{1}{|c|}{}       & \multicolumn{1}{c}{$\mu$} & \multicolumn{1}{c|}{$\sigma$} & \multicolumn{1}{c}{$\mu$} & \multicolumn{1}{c|}{$\sigma$} & \multicolumn{1}{c}{$\mu$} & \multicolumn{1}{c|}{$\sigma$} \\ \hline
		0.1                      & 2.8593			    &1.7050		   	& 2.8301		    &1.6989		   & 2.8007 			   &1.6927\\
    		0.3			&2.9634			    &1.6881			&2.8690			&1.6686			&2.7766			    &1.6496\\
    		0.5			&3.0714			    &1.6573			 &2.9110			 &1.6260			 &2.7558			    &1.5944\\
    		0.7			&3.1754			    &1.5873			& 2.9597 			  &1.5469			 &2.7557			    &1.5071\\
    		0.9			&3.2582			    &1.4922			 &3.0057			  &1.4457			  &2.7672			    &1.3991                       \\ \hline
	\end{tabular}
	\caption{Values of the parameters $\mu$ and $\sigma$ when the truncated normal distribution (\ref{eq:truncDist}) subject to the constraint (\ref{eq:constraint}) was fit  to the direct simulation (continuous) model and to  the Markov chain (surrogate) model.}
	\label{tab:PDF_fits}
\end{table}

\begin{figure}[h!]  
\bigskip
\begin{center}
\mbox{ 
\includegraphics[scale=0.26]{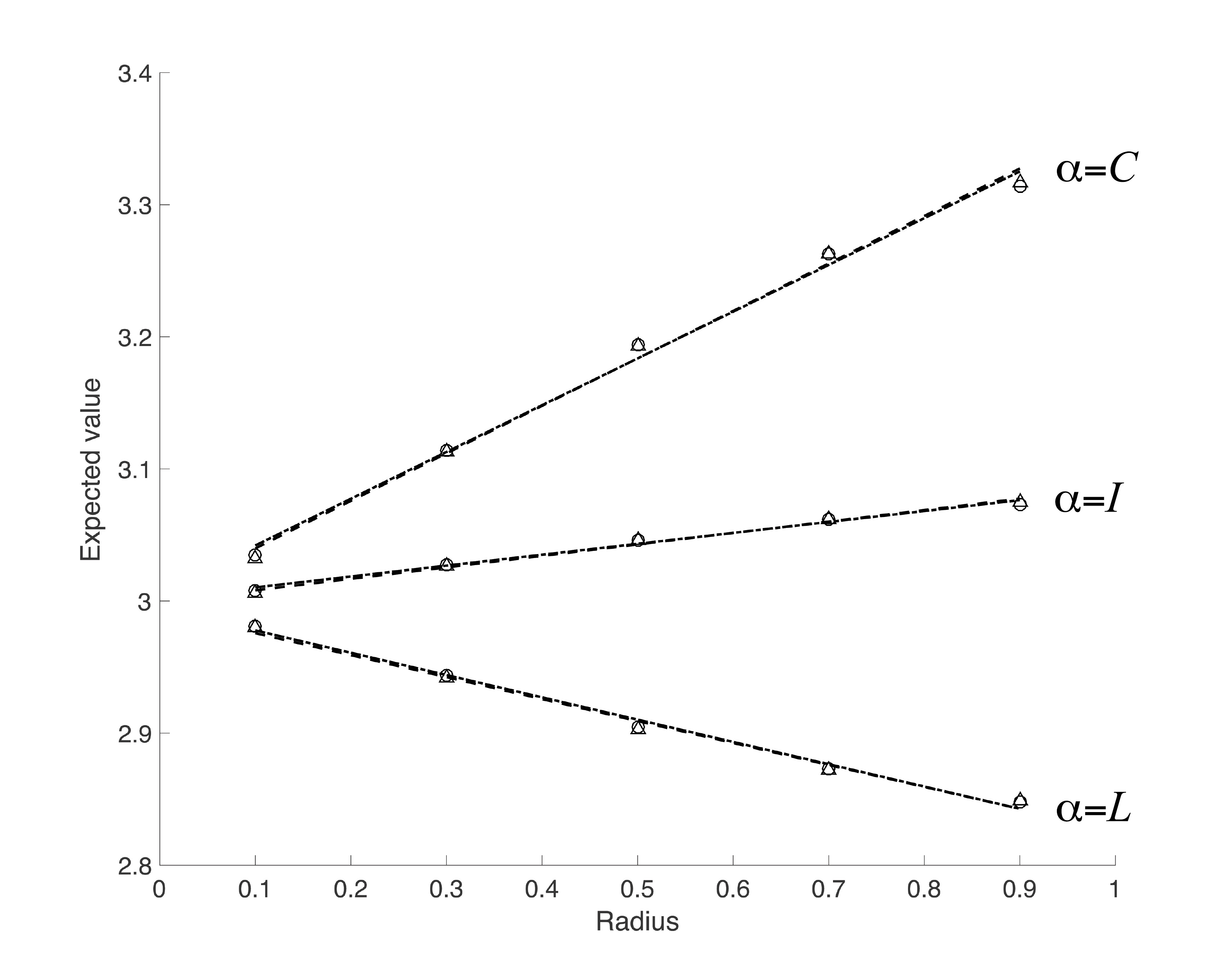}
}
\end{center}
\caption{Linear regression fits for the relationship between the average number of particles per subdomain $\bar{\eta}_\alpha, \alpha=L,I,C$ and the particle radius $R$, delineated by subomain type.}
\label{linreg}
\end{figure}

\begin{table}[h!]
\centering
\begin{tabular}{|l|c|c|c|c|c|c|}
\hline 
& \multicolumn{3}{|c|}{Continuous} & \multicolumn{3}{|c|}{Markov Chain} \\
\hline
  & Slope & Intercept & $R^2$ & Slope & Intercept & $R^2$ \\
\hline
Center & 0.3539 & 3.0067 & 0.9929 & 0.3595 & 3.0039 & 0.9939 \\
Onewall & 0.0826 & 3.0021 & 0.9895 & 0.0868 & 2.9995 & 0.9921 \\
Corner & -0.1688 & 2.9946 &  0.9928 & -0.1663& 2.9923 &  0.9903\\
\hline
\end{tabular}
\caption{Slope, intercept and coefficient of determination ($R^2$) for a linear regression analysis of the relationship between the  average number of particles per subdomain $\bar{\eta}_\alpha, \alpha=L,I,C$ and the particle radius $R$ for the continuous and the Markov chain models.}
\label{tab:linreg}
\end{table}

The  bounds for the range of our random variable, which is the number of particles per subdomain (quantity of interest),  were prescribed as  $[a,b]=[0,N_s]$, recalling that $N_s$ is the maximum number of states per subdomain  in the Markov chain model.  An appropriate choice for $N_s$ was  determined by increasing its value until curve fits of the truncated normal  distribution  to the stationary distributions in the Markov chain model $\vec{\pi}_\alpha$  stabilized their shape. It should be noted that the occurrence of states becomes rarer as the number of particles per subdomain increases.  The  small probability values in $\vec{\pi}_\alpha$  at these larger values make the process of fitting the truncated normal distribution quite sensitive to the method of curve-fitting.  To address this challenge, the distributions were fit to the data using constrained optimization via the ``fmincon'' function in MATLAB (using the SQP option) with incorporation of  the following constraint:
\begin{equation}
\mu + \frac{\phi\left(\frac{a-\mu}{\sigma} \right) - \phi\left(\frac{b-\mu}{\sigma} \right)}
{\Phi \left(\frac{b - \mu}{\sigma}\right) - \Phi \left( \frac{a - \mu}{\sigma}\right)} 
- \sum_{j=0}^{N_s} j \pi_{\alpha}^j  = 0,\;\;\;\alpha=L,I,C.
\label{eq:constraint}
\end{equation}
Equation (\ref{eq:constraint}), with $a=0$ and $b=N_s$, expresses the condition that the mean of the truncated normal distribution (\ref{eq:PDFs}) should equal the expected value for the Markov chain model when $N_s$ states are included in the surrogate model.

An illustration of the process of refining $N_s$, including fits to the stationary distributions based on (\ref{eq:truncDist})-(\ref{eq:constraint}), is shown in Fig.~\ref{fig:TN_states}.  Based on this process, all subsequent results  presented herein employed a value of $N_s=13$.    
The resulting Markov chain stationary distributions ($\vec{\pi}_\alpha$ from (\ref{eq:statdist})), corresponding continuous model data, and fits to both using the truncated normal distribution (\ref{eq:truncDist}) were determined for each subdomain type ($\alpha=L,I,C$) and with varying particle radius $R$ in the range $0.1$-$0.9$  (Fig.~\ref{fig:TNall}). Excellent agreement is  observed between stationary distributions calculated using the continuous model and the (surrogate) Markov chain model.  
The corresponding values of the estimated parameters $\mu$ and $\sigma$ for the fits shown in Fig.~\ref{fig:TNall} (see Table \ref{tab:PDF_fits}) demonstrate excellent agreement between the continuous and surrogate (MC) model for all cases shown in Fig.~\ref{fig:TNall}.  Effects of increasing the particle radius $R$ for each of the three subdomain types resulted in significant shifts to the right in the distributions for the Center ($\alpha=C$) and One-wall ($\alpha=I$) subdomains, and much less pronounced variation for the Corner ($\alpha=L$) subdomains (Fig.~\ref{fig:TNR}).

Lastly, a regression analysis was performed to determine empirical relations between the mean value of the number of particles per subdomain ($\bar{\eta}_{\alpha}$) and the particle radius $R$ for each of the three subdomain types ($\alpha=L,I,C$) (Fig.~\ref{linreg}, Table \ref{tab:linreg}).   Results demonstrated that linear regression provided excellent fits ($R^2>0.98$) for all three subdomain types.  In the Center ($\alpha=C$) and Corner ($\alpha=L$) cases, the slope and intercept in the regression fits for the continuous model and the (surrogate) Markov chain model agreed to within $1.6\%$ relative error.  In the case of the One-wall subdomains ($\alpha=I$), the relative error was larger at a value of $5.1\%$, but it is noted that the slopes for this case are closer to zero as compared to the other two subdomain cases.   The loss of available area as the particle radius $R$ is increased is reflected in the negative slope for the Corner subdomains where the area available to be occupied by particles is significantly reduced as  $R$ increases. This is countermanded by the relatively large positive slope for the Center subdomain for which no available area is lost with increasing particle radius.  Conversely, as $R\rightarrow 0$, all three regression lines appear to converge  to a  value of $3.00$ particles per subdomain, as evidenced by the intercept values  in Table \ref{tab:linreg}.

\section{Discussion and Conclusions}
\label{sec:conclusions}

This study investigated  a two-dimensional system of  circular particles interacting, via perfectly elastic collisions, with each other and with the four walls of a square domain. By partitioning the simulation domain into 9 equal square subdomains, statistical properties of the system were delineated based on three subdomain types with differing  geometric features.  By taking the number of particles per subdomain as the quantity of interest, a surrogate model was formulated based on Markov chains.  The states in the Markov chain model were the number of particles per subdomain, in each of the three subdomain categories, with transitions occurring between adjacent subdomains.  

Excellent agreement between the directly simulated continuous model and the surrogate (Markov) model was achieved by tracking 14 states, i.e.~significantly less than the total of 27 particles.  Statistics for the quantity of interest indicated that a truncated normal distribution was well-suited to capturing statistical properties in estimates of the number of particles per subdomain type.  Expected values were found to vary linearly with increasing particle radius $R$, and increased with $R$ for subdomains with zero walls or one wall, but decreased for subdomains with two walls.  Results of this type can be used to estimate the aggregate density of particles, via direct simulations in a representative domain, since it is unclear that such measures can be predicted a priori, except in the limit $R\rightarrow 0$. Indeed, the uniform density in this limiting case is consistent with Brownian motion (diffusion) on a square with insulated boundaries, yet the density profile as $R$ increases is difficult to predict analytically.    Beyond expected values, the modeling approach presented herein has the advantage  that it includes an estimate of uncertainty via the use of truncated normal distributions that were demonstrated to be well-suited for analyzing the quantity of interest.

The approach employed in this study may have potential application in other systems where aggregate properties are unknown but can be investigated via direct simulations of discrete entities on a representative domain.  Success will depend on the ability to accurately and efficiently compute solutions using the direct model on time scales and for enough realizations to exhibit stationary statistical properties for the quantities of interest.  More specifically, development of a surrogate model using Markov chains will require that the quantities of interest have a readily identified set of states and that the notion of transitions between these states can be easily defined. When states are continuous rather than discrete quantities, it is possible that a set of discrete states can still be defined by binning the random variable into adjoining sub-ranges of the continuous variable.  The viability of this approach for such systems needs further investigation. When accurate Markov chain surrogate models can developed, they also have the potential to serve as useful tools for accelerating portions of simulations for complex systems while also  providing a quantitative framework for uncertainty quantification.  

Overall, the methods and approach developed in this study for a simpler two-dimensional system may have potential utility in multiscale modeling of more complex systems with discrete entities (e.g. particles, polymers or biological cells) exhibiting non-trivial dynamical interactions that can be directly and efficiently simulated on a representative domain.

\section{Acknowledgements}
Work supported by the National Science Foundation via grants DMS-1638521 and a NSF Graduate Research Fellowship.

\section{Competing Interests Declaration}
The author(s) declare none.

\bibliographystyle{siamplain}

\end{document}